\documentstyle[preprint,eqsecnum,aps,epsfig]{revtex}

\begin{document}
\preprint{UNIGRAZ-UTP 15-10-99}
\title{Hard exclusive photoproduction of $\Phi$ mesons}
\author{C. F. Berger\cite{berger}}
\address{Department of Physics and Astronomy, SUNY at Stony Brook, Stony Brook, 
NY 11794-3800, USA}
\author{W. Schweiger\cite{schweiger}}
\address{Institut f\"ur Theoretische Physik, Universit\"at Graz, Universit\"atsplatz 5, 
A-8010 Graz, Austria}
\date{October 27, 1999}

\maketitle
\newpage
\begin{abstract} 
We calculate the differential cross section and
single-polarization observables for the reaction $\gamma \, p \, \rightarrow  \, \Phi \,  p$ within 
perturbative QCD, 
treating the proton as a
quark-diquark system. The phenomenological couplings of gauge bosons to
(spatially extended) diquarks and the quark-diquark distribution
amplitude of the proton are adopted from previous investigations of baryon
form factors and two-photon processes. Going beyond leading order, we take
into account hadron-mass effects by means of a systematic expansion in the
small parameter (hadron mass/ photon energy). With the $\Phi$-meson
distribution amplitude taken from the literature our  predictions for the differential cross section 
at
$\vert t \vert \agt 4 \text{GeV}^2$ seem to provide a reasonable extrapolation of
the low-$t$ data and are also comparable in magnitude with the results of
a two-gluon exchange model in which the gluons are considered as a
remnant of the pomeron. For momentum transfers of a few GeV hadron-mass
effects appear still to be sizeable.
\end{abstract}


\narrowtext

\section{Introduction}
\label{sec:intro}
In the course of the last few years exclusive photo- and electroproduction of neutral vector 
mesons ($\rho$,
$\omega$, $\Phi$, $J/\Psi$) at intermediate and high energies has attained increased theoretical 
interest, 
which has been stimulated by corresponding  experimental efforts at CEBAF~\cite{Au93} and 
HERA~\cite{Cri97}.
The aim of the HERA experiments is a more fundamental understanding of Pomeron 
phenomenology in terms
of QCD. Vector-meson dominance in combination with Pomeron exchange proves to be very 
successful in
describing diffractive photo- and electroproduction of neutral vector mesons, provided the 
energy is the only
large scale in the process~\cite{DL95}. In their approach Donnachie and 
Landshoff~\cite{DoLa87} assume that
the pomeron couples directly to on-shell quarks in the vector meson and the proton, 
respectively.
A QCD-inspired version of Pomeron exchange, in which the Pomeron is replaced by two non-
perturbative (abelian)
gluons, goes back to the same authors~\cite{DoLa89} and is usually termed a \lq\lq soft 
Pomeron\rq\rq. In
the limit $t \rightarrow 0$ the soft-pomeron exchange becomes very similar to the usual 
Pomeron exchange.
This can be used to interpret the form factor at the quark-Pomeron vertex~\cite{DoLa89}. 
However, if in
addition to the energy another scale, like the photon virtuality or the mass of the vector meson, 
becomes
large perturbative QCD effects tend to enter the game. For large photon virtuality (and small 
values of
Mandelstam $t$) perturbative QCD becomes applicable due to a factorization theorem which has 
been
proved in Ref.~\cite{CFS97}. The three building blocks of the factorization formula are the 
hard
photon-parton scattering amplitude, the distribution amplitude of the vector meson and finally a 
skewed
gluon distribution of the proton.  As a  link between hard inclusive and exclusive reactions 
which
generalizes the ordinary parton distributions, skewed parton distributions have recently attracted 
a great
amount of interest (see, e.g., Ref.~\cite{DFJK99} and references therein). Electroproduction 
with highly
virtual photons thus offers a possibility to study such skewed parton distributions. Within this 
perturbative
formalism for vector-meson electroproduction with highly virtual photons the two-gluon aspect 
of the pomeron
emerges automatically in a form which is related to the skewed gluon distribution in the proton.

One of the aims of photoproduction experiments at CEBAF is, on the other hand, to understand 
the
transition from non-perturbative to perturbative production mechanisms. In the case of 
photoproduction the
transverse momentum transfer  $p_\perp^2 \approx t u / s$ has to be large in order that 
perturbative QCD becomes
applicable. The pertinent factorization formula, which in general holds for arbitrary exclusive 
hadronic
processes at large transverse momenta, is the result of a leading-order perturbative analysis. It 
represents 
a hadronic amplitude as a convolution integral of a hard-scattering amplitude with distribution 
amplitudes
(DAs)~\cite{BL89}. The process-dependent hard-scattering amplitude is perturbatively 
calculable and
represents the scattering of hadronic constituents in collinear approximation. The process-
independent DAs
contain the non-perturbative bound-state dynamics of the hadronic constituents. There is no 
doubt that this 
hard-scattering approach (HSA) gives the correct description of exclusive hadronic processes in 
the limit of
asymptotically large momentum transfer. But there is increasing evidence that the pure 
perturbative
contribution is not the dominant one at experimentally accessible momentum transfers. A 
reasonable
reproduction of the existing exclusive large-$t$ data is only achieved with very endpoint-
concentrated
hadron DAs, like those proposed by Chernyak and Zhitnitsky on the basis of QCD sum 
rules~\cite{CZ84}. Such
DAs, however, occur to be problematic for a perturbative calculation since they enhance
contributions from the soft end-point regions in the convolution integral~\cite{IL89}. Some 
improvement with
regard to the consistency of the perturbative calculation can be achieved by applying a modified
factorization scheme in which transverse-momentum effects and Sudakov suppressions are 
included. But this
leads to a substantial reduction of the perturbative contribution~\cite{BJKBS95}.  For the pion 
an
end-point concentrated DA meanwhile seems even to be ruled out by the recent CLEO 
data~\cite{CLEO98} for
the $\pi \gamma$-transition form factor, since it provides a much too large perturbative result. 
On the
other hand, it has been shown recently for electromagnetic nucleon form factors~\cite{BK96} 
and for
wide-angle Compton scattering~\cite{DFJK99,Ra98} off protons that the existing data can be 
saturated by
soft overlap contributions. Furthermore, these contributions can be modeled by a nucleon light-
cone wave
function which gives rise to a distribution amplitude close to the asymptotic one. This, 
however, implies
that the perturbative contribution becomes small.

The approach which we are going to apply for hard exclusive photoproduction of $\Phi$ 
mesons is still based
on the HSA but includes also a certain amount of soft physics which is modeled by means of 
diquarks. Diquarks
may be considered as an effective way to account for quark-quark correlations in baryon wave 
functions. Part
of the soft overlap contributions are then absorbed into phenomenologically parameterized 
diquark form factors
which occur at the gauge-boson diquark vertices. This HSA-based diquark model has already 
been applied
successfully to other photon-induced hadronic reactions like magnetic and electric baryon form 
factors in the
space- \cite{Ja93} and time-like region \cite{Kro93}, real and virtual Compton scattering 
\cite{Kro95,Be97},
two-photon annihilation into baryon-antibaryon \cite{Kro93,Be97}, and photoproduction of 
$K$-mesons~\cite{KSPS97}.
Further applications of the diquark model include the charmonium decay $\eta_{\text{c}} 
\rightarrow p
\bar{p}$ \cite{Kro93} and the calculation of Landshoff contributions in elastic proton-proton
scattering \cite{Ja94}. The purpose of the present investigation is to extend this foregoing work 
to $\gamma\, p\, \rightarrow\, \Phi\,
p$. The reason why we concentrate on the $\Phi$ channel is that it is cleaner and simpler
than photoproduction of $\rho$ or $\omega$ mesons. Within the HSA the valence Fock state of 
the $\Phi$, i.e.
the $s$-$\bar{s}$ state, can only be produced via the exchange of two gluons between the 
$\Phi$ and
the proton. Quark exchange could only happen via the knockout of an $s$-$\bar{s}$ pair from 
the strange-quark
sea in the nucleon which is, however, power suppressed within the HSA. 

In the following section we start with a short outline of the diquark model, show how Feynman 
diagrams are
properly grouped according to their propagator singularities (Sec.~\ref{sec:elamp}), sketch how 
we deal with
the propagator singularities numerically (Sec.~\ref{sec:propnum}), and present our treatment of 
hadron-mass
effects (Sec.~\ref{sec:mass}). Section~\ref{sec:helamp} summarizes how photoproduction of 
vector mesons is
described in terms of helicity amplitudes. Furthermore, the fixed-angle scaling behavior of these 
helicity
amplitudes is discussed and related to the treatment of mass effects. Analytical expressions for 
the hard
part of the hadron-helicity-conserving amplitudes are also given in this section. Our predictions 
for
photoproduction observables along with predictions from Pomeron and soft-Pomeron exchange 
are presented in
Sec.~\ref{sec:result}. At the end of this section we comment on the prospects of extending our 
calculation to
photoproduction of $J/\Psi$ mesons. Finally, a summary and concluding remarks can be found 
in
Sec.~\ref{sec:sum}.

\section{Formalism}

\subsection{The hard-scattering formalism with diquarks}
\label{sec:diquark}
As already mentioned before, within the HSA an exclusive scattering amplitude at large 
momentum transfer is expressed 
as a convolution integral of a hard-scattering amplitude with hadron distribution 
amplitudes~\cite{BL89}. For
the photoproduction reaction $\gamma \, p\, \rightarrow\, \Phi \, p$, which we are interested in, 
this
integral takes on the particular form
\widetext
\begin{equation}
M_{\{\lambda\}}(\hat{s},\hat{t}) \! = \! \int_0^1 \! \! dx_1 dy_1 dz_1
{\phi^\Phi}^{\dagger}(z_1,\tilde{p}_{\! \perp})
{\phi^p}^{\dagger}(y_1,\tilde{p}_{\! \perp})
\widehat{T}_{\{\lambda\}}(x_1,y_1,z_1;\hat{s},\hat{t})
\phi^p(x_1,\tilde{p}_{\! \perp}) \, . \label{convol}
\end{equation}
\narrowtext\noindent
The distribution amplitudes $\phi^H$ are probability amplitudes for
finding the valence Fock state in the hadron $H$ with the constituents carrying certain
fractions of the momentum of their parent hadron and being  collinear up to a
factorization scale $\tilde{p}_{\perp}$. In our model the valence Fock state of an ordinary
baryon is assumed to consist of a quark ($q$) and a diquark ($D$). We fix our notation in
such a way that the momentum fraction appearing in the argument of $\phi^{H}$ is
carried by the quark -- with the momentum fraction of the other constituent (either
diquark or antiquark) it sums up to 1 (cf. Fig.~\ref{kinem}). For our actual
calculations the (logarithmic) $\tilde{p}_{\perp}$ dependence of the DAs is neglected
throughout since it is of minor importance in the restricted energy and
momentum-transfer range we are  interested in. The hard scattering amplitude
$\widehat{T}_{\{\lambda\}}$ is calculated perturbatively.  Intrinsic transverse momenta of the
constituents are thereby neglected, i.e., one assumes that the momenta of the constituents are
collinear to those of their parent hadron.
$\widehat{T}_{\{\lambda\}}$ consists in our particular case of all possible tree diagrams
contributing to the elementary scattering process  $\gamma\,q\, D\, \rightarrow\, Q \, \bar{Q}\,
q \, D$. Eight of the, altogether, sixteen diagrams which contribute to photoproduction of
(heavy) quarkonia are depicted in Fig.~\ref{feyn}. The subscript ${\{\lambda\}}$ of
$\widehat{T}$ represents a set of possible photon, proton and vector-meson helicities. For our 
purposes
it appears to be more convenient to express the analytical results in terms of massless 
Mandelstam variables
$\hat{s}$, $\hat{t}$, and $\hat{u}$ than in terms of the usual (massive) ones. A detailed 
account of our
treatment of hadron-mass effects is given in Sec.~\ref{sec:mass} below.

The diquark model, as applied in Refs.~\cite{Ja93}-\cite{KSPS97}, comprises scalar ($S$) as
well as axial-vector ($V$) diquarks. The dynamics of $S$ and $V$ diquarks is determined by 
their
coupling to photons and gluons. The expressions for the photon-diquark and gluon-diquark 
vertices
correspond (almost) to the most general form for the (parity and time-reversal invariant) 
coupling of
a spin-1 gauge boson to a spin-0 or a spin-1 particle, respectively~\cite{Ja93}. Thereby $V$
diquarks are allowed to possess an anomalous (chromo)magnetic moment $\kappa_V$. In 
applications
of the model Feynman graphs are calculated first with the corresponding Feynman rules for
point-like diquarks. As a second step, Feynman graphs with $(n-2)$ gauge bosons coupling to 
the
diquark are combined in a group which we call \lq\lq n-point contribution\rq\rq.\ Finally 
the composite nature of diquarks is taken into account by multiplying each n-point contribution 
with
phenomenological vertex functions, the diquark form factors. The particular choice      
\begin{eqnarray} 
F_S^{(3)} (Q^2) &=& \; \delta_S \, \frac{Q_S^2 }{ (Q_S^2 +
Q^2)}\, ,  \\
F_V^{(3)} (Q^2) &=& \; \delta_V \left( \frac{Q_V^2 }{ Q_V^2 + Q^2}\right)^2 \, ,
\end{eqnarray}  
for 3-point contributions and 
\begin{eqnarray}
F_S^{(n)} &=& a_S F_S^{(3)} (Q^2) \, ,  \\
F_V^{(n)} &=& a_V F_V^{(3)} (Q^2) \left( \frac{Q_V^2 }{ Q_V^2 + Q^2 }\right)^{(n-3)} \, 
,
\end{eqnarray} 
for n-point contributions ($n \geq 4$) ensures that in the limit $Q^2 \rightarrow
\infty$ the scaling behavior of the diquark model goes over in that of the pure
quark HSP. The factor $\delta_{S (V)}  =  \alpha_s (Q^2) / \alpha_s (Q^2_{S (V)})$
($\delta_{S (V)} = 1$ for $Q^2 \leq  Q^2_{S (V)}$) provides the correct powers of
$\alpha_s (Q^2)$ for asymptotically large $Q^2$. For the running coupling $\alpha_s$ the
one-loop result $\alpha_s = 12 \pi / 25 \ln (Q^2 / \Lambda_{QCD}^2 )$ 
is used with $\Lambda_{QCD} = 200 {\rm \text{MeV}}$. In addition $\alpha_S$ is restricted
to be smaller than $0.5$. $a_S$ and $a_V$ are strength parameters which allow for the
possibility of diquark excitation and break-up in intermediate states where diquarks can
be far off-shell. 

Having sketched how diquarks are treated perturbatively, a few words about the choice
of the DAs (which incorporate the bound-state dynamics of the hadronic constituents) and the
spin-flavor wave functions of the proton and the $\Phi$ are in order. 
If one assumes zero relative orbital angular momentum between quark and diquark
and takes advantage of the collinear configuration ($p_q = x_1 p_B$ and
$p_D = x_2 p_B = (1 - x_1) p_B$) the valence Fock-state wave function of an octet baryon B 
may be
written as
\widetext
\begin{equation}
\Psi^B ( p_B; \lambda) =
f_S^B
\phi_S^B (x_1) \, \chi^B_S \,
u( p_B,\lambda)
 + f_V^B \, \phi_V^B (x_1) \,
 \chi^B_V \, \frac{ 1 }{ \sqrt{3}} \, (\gamma^\alpha +
\frac{p_B^\alpha }{ m_B }) \, \gamma_5 u(
p_B,\lambda )
\, .
\label{wfcov}
\end{equation}
\narrowtext\noindent
The two terms in Eq.~(\ref{wfcov}) represent states consisting of a
quark \hbox{and} either a scalar or a vector diquark. The pleasant feature of the covariant
wave-function representation, Eq.~(\ref{wfcov}), is that it contains, besides $x_1$ and
$\alpha$ (the Lorentz index of the vector-diquark polarization vector), only baryonic
quantities (momentum $p_B$, helicity $\lambda$, baryon mass $m_B$).
Assuming an SU(6)-like spin-flavor dependence, the flavor functions $\chi_D^p$ for a
proton take on the form
\begin{equation}
\chi_S^p = u S_{[u,d]} \, , \;\;\;
\chi_V^p = \phantom{-} [u V_{\{u,d\}}
-\sqrt{2} d
V_{\{u,u\}}] / \sqrt{3} \, .
\label{flavp}
\end{equation}
In Eq.~(\ref{wfcov}) the color part of the quark-diquark wave function has been neglected. 
Since the 
quark-diquark state has to be a color singlet the diquark has to be in a color antitriplet state (like 
an
antiquark) Thus the quark-diquark color wave function is simply $\psi_{qD}^{\text{color}} =
(1/\sqrt{3}) \sum_{a=1}^3 \delta_{a \bar{a}}$.

Like the baryon wave function, Eq.~(\ref{wfcov}), also the $Q$-$\bar{Q}$ wave function of a 
vector 
meson $M$ can be represented in a covariant way
\begin{equation}
\Psi^{M} ( p_{M}; \lambda) =
- f^{M}  \, \phi^{M} (z_1,\lambda) \, \chi^{M} \,
\frac{1}{ \sqrt{2}}(\not{\!p}_{M} +  m_{M} )
\not{\!\epsilon}(\lambda)
\, , \label{wfcovv}
\end{equation}
with $\epsilon(\lambda)$ denoting the polarization vector of the vector meson.
The flavor function of the $\Phi$ is $\chi^{\Phi} = s \bar{s}$.

In Refs.~\cite{Ja93}-\cite{KSPS97} a quark-diquark DA of the form ($c_1 = c_2 = 0$ for
$S$ diquarks)
\widetext 
\begin{equation}
\phi_{D}^{B}(x_1)  = N^B x_1 (1-x_1)^3 (1 + c_1 x_1 + c_2 x_1^2)
\exp \left[ - b^2 \left( \frac{m_{q}^2}{x_1} + \frac{m_{D}^2 }{
(1-x_1)} \right) \right] \, , 
\label{DAp}
\end{equation}
\narrowtext\noindent
turned out to be quite appropriate for octet baryons $B$. The origin of the phenomenological 
baryon DA,
Eq.~(\ref{DAp}), is a non-relativistic harmonic-oscillator wave function (cf. 
Ref.~\cite{Hua89}). Therefore
the masses appearing in the exponential have to be considered as constituent masses (330 MeV 
for light
quarks, 580 MeV for light diquarks, strange quarks are 150 MeV heavier than light quarks). 
The oscillator
parameter $b^2 = 0.248$ GeV$^{-2}$ is chosen in such a way that the full wave function gives 
rise to a value
of $600$ MeV for the mean intrinsic transverse momentum of quarks inside a nucleon. The 
exponential does
not only provide a flavor dependence, but also suppresses the end-point regions $x_1,\, y_1\, 
\rightarrow
0,\, 1$ in the convolution integral, Eq.~(\ref{convol}). One thus avoids to pick up sizable 
contributions from
the soft end-point regions and results become less dependent on details of the baryon DA. In the 
actual
data fitting the exponential plays only a minor role. Therefore, the constituent masses and the 
oscillator
parameter have not been considered as free parameters but have been fixed in advance. We 
stress that the constituent
masses do not appear in the hard scattering amplitudes (cf. Sec.~\ref{sec:mass} below).

A model for the $\Phi$ meson DA which fulfills QCD sum-rule constraints has been proposed
by Benayoun and Chernyak~\cite{BC90}. Since we also want to investigate to which extent our 
calculations for
$\Phi$ production may be generalized to  $J/\Psi$ production we modify this DA by attaching
an exponential factor which provides a flavor dependence. The effect of this factor is negligible 
for
$\Phi$s, but it becomes crucial for $J/\Psi$s.  The form of the exponential factor is inspired by a 
simple
relativistic treatment of heavy mesons~\cite{BSW85} and provides a flavor dependence in
accordance with heavy-quark effective theory. For longitudinally polarized
$\Phi$s and $J/\Psi$s the DA reads ($M = \Phi,  J/\Psi$)
\begin{equation}
\phi^M (z_1,\lambda_\Phi=0) = N^M_{\lambda_\Phi=0} z_1 (1 - z_1) \exp \left[ - \tilde{b}^2 
m_{M}^2 (z_1 - 0.5)^2
\right]\, .
\label{DABC}
\end{equation}
The DA for transversally polarized $\Phi$s and $J/\Psi$s contains an additional polynomial
in $z_1$ which enhances the maximum at $z_1=0.5$ and makes it narrower (cf. 
Ref.~\cite{BC90})
\widetext
\begin{eqnarray} 
\phi^M (z_1,\lambda_\Phi= \pm 1) & = & N^M_{\lambda_\Phi=\pm 1} z_1 (1 - z_1) 
\exp \left[ - \tilde{b}^2 m_{M}^2 (z_1 - 0.5)^2 \right] \nonumber \\ & & \times
(5225.2\, z_1^6 (1 - z_1)^6 + 0.39\, z_1 (1 - z_1))
\, .
\label{DABCl}
\end{eqnarray}
\narrowtext\noindent
For the oscillator parameter $\tilde{b}$ we take a value of $0.97$~GeV$^{-1}$ in
accordance with estimates for the radius of the $J/\Psi$ meson~\cite{FK97}. 

The usual normalization 
condition, $\int_0^1 dx \phi^{H}(x) = 1$, fixes the constants $N^B$ and 
$N^M_{\lambda_\Phi}$ in
Eqs.~(\ref{DAp}), (\ref{DABC}), and (\ref{DABCl}). The quantity $f^{M}$ showing up
in Eq.~(\ref{wfcovv}) is related to the experimentally determinable decay constant
of the meson $M$. In case of the $\Phi$ we get from the radiative decay width of the $\Phi 
\rightarrow
e^+ e^-$ decay $f^{\Phi} = f^{\Phi}_{\text{decay}}/{2 \sqrt{6}} = 46.95$~MeV. The
analogous constants $f_S^B$ and $f_V^B$ for the $q$-$D$ DAs of baryon $B$ are
free parameters of the diquark model. They are, in principle, determined by the probability
to find the $q$-$D$ state ($D = S, V$) in the baryon $B$ and by the transverse-momentum
dependence of the corresponding wave function.

Our present study is performed with the set of diquark-model parameters found in 
Ref.~\cite{Ja93} by
fitting elastic electron-nucleon scattering data. The numerical values of the parameters are:
\begin{eqnarray}
f_S &=& 73.85 \hbox{ MeV}, \; Q_S^2 = 3.22 \hbox{ GeV}^2, \; a_S = 0.15,  \nonumber \\ 
f_V &=& 127.7 \hbox{ MeV}, \; Q_V^2 = 1.50 \hbox{ GeV}^2, \; a_V = 0.05,  \nonumber \\
\kappa_V &=& 1.39,\qquad  c_1 = 5.8, \qquad c_2 = -12.5 \, .  \label{param}
\end{eqnarray}
For further details of the diquark model we also refer to the paper of Jakob et al.~\cite{Ja93}.

\subsection{The general structure of the \\ hard-scattering amplitude}
\label{sec:elamp}
The introduction of diquark form factors already requires a decomposition of the hard-scattering 
amplitude
$\widehat{T}_{\{ \lambda \}}$ into 3- and 4-point contributions
\begin{eqnarray}
\widehat{T}_{\{\lambda\}}(x_1,y_1,z_1;\hat{s},\hat{t}) & = & 
e_s \widehat{T}_{\{\lambda\}}^{(Q,S)}(x_1,y_1,z_1;\hat{s},\hat{t}) \nonumber \\ &+&
e_u \widehat{T}_{\{\lambda\}}^{(q,S)}(x_1,y_1,z_1;\hat{s},\hat{t}) \nonumber \\ &+&
e_{S_{[u,d]}} \widehat{T}_{\{\lambda\}}^{(S)}(x_1,y_1,z_1;\hat{s},\hat{t}) \nonumber \\ 
&+&
e_s \widehat{T}_{\{\lambda\}}^{(Q,V)}(x_1,y_1,z_1;\hat{s},\hat{t}) 
\, .
\label{Tn}
\end{eqnarray}
$e_q$ and $e_{S_{[u,d]}}$ are the electric charges of the quark $q$ and the S diquark 
$S_{[u,d]}$ in units of $e_0$, 
respectively.
The superscripts indicate whether the photon couples to the heavy quark $Q$ in the meson, or 
the light quark
$q$ in the proton with the diquark $D = S,V$ acting as spectator (3-point contributions), or 
whether the photon
couples to the $S$ or $V$ diquark in the proton (4-point contributions). Note that the only $V$-
diquark
contribution is $\widehat{T}_{\{\lambda\}}^{(Q,V)}$. The reason is that the charge-flavor 
factor in front of
$\widehat{T}_{\{\lambda\}}^{(q,V)}$ vanishes and $\widehat{T}_{\{\lambda\}}^{(V)}$ 
represents already a mass
correction of second order, which is neglected in our calculation. For the numerical treatment of 
the
convolution integral, Eq.~(\ref{convol}), it is advantageous to further split the $n$-point 
contributions into
two parts which differ by their propagator singularities:
\widetext
\begin{eqnarray}
\widehat{T}_{\{\lambda\}}^{(Q,D)}(x_1,y_1,z_1;\hat{s},\hat{t})
& = & \left[ \frac{f_{\{\lambda\}}^{(Q,D)}(x_1,y_1,z_1;\hat{s},\hat{t})}
           {(\hat{q}_2^2 + i \epsilon) (\hat{q}_5^2 + i \epsilon^\prime)}
  +   \frac{g_{\{\lambda\}}^{(Q,D)}(x_1,y_1,z_1;\hat{s},\hat{t})} 
           {(\hat{q}_3^2 + i \epsilon) (\hat{q}_4^2 + i \epsilon^\prime)} \right] 
F_D^{(3)}(-x_2 y_2 \hat{t})
\, , \nonumber \\
\widehat{T}_{\{\lambda\}}^{(q,D)}(x_1,y_1,z_1;\hat{s},\hat{t})
& = & \left[ \frac{f_{\{\lambda\}}^{(q,D)}(x_1,y_1,z_1;\hat{s},\hat{t})}
           {(\hat{g}_1^2 + i \epsilon) (\hat{q}_2^2 + i \epsilon^\prime)}
  +   \frac{g_{\{\lambda\}}^{(q,D)}(x_1,y_1,z_1;\hat{s},\hat{t})} 
           {(\hat{g}_1^2 + i \epsilon) (\hat{q}_3^2 + i \epsilon^\prime)} \right] 
F_D^{(3)}(-x_2 y_2 \hat{t})
\, , \nonumber \\
\widehat{T}_{\{\lambda\}}^{(D)}(x_1,y_1,z_1;\hat{s},\hat{t})
& = & \left[ \frac{f_{\{\lambda\}}^{(D)}(x_1,y_1,z_1;\hat{s},\hat{t})}
           {(\hat{g}_2^2 + i \epsilon) (\hat{q}_5^2 + i \epsilon^\prime)}
  +   \frac{g_{\{\lambda\}}^{(D)}(x_1,y_1,z_1;\hat{s},\hat{t})} 
           {(\hat{g}_2^2 + i \epsilon) (\hat{q}_4^2 + i \epsilon^\prime)} \right] 
F_D^{(4)}(-x_2 y_2 \hat{t}) \, .
\label{Tfg}
\end{eqnarray}
\narrowtext\noindent
The functions $f$ and $g$ correspond to gauge-invariant subgroups of the 16 Feynman 
diagrams which
contribute to the hard-scattering amplitude $\widehat{T}_{\{ \lambda \}}$ (cf. 
Fig.~\ref{feyn}). The
diagrams entering the $g$s are related to those entering the $f$s by interchange of the two 
gluons. The
$q_i^{-2}$ and $g_i^{-2}$ denote just those quark and gluon propagators which can go on-
shell when integrating
over $x_1$, $y_1$, and $z_1$. Explicitly, the propagator denominators read:
\begin{eqnarray}
\hat{q}_2^2 & = y_2 z_2 \hat{s} + x_2 y_2 \hat{t} + x_2 z_2 \hat{u} \, , \quad
\hat{q}_3^2 & = y_2 z_1 \hat{s} + x_2 y_2 \hat{t} + x_2 z_1 \hat{u} \, , \nonumber \\
\hat{q}_4^2 & = y_1 z_2 \hat{s} + x_1 y_1 \hat{t} + x_1 z_2 \hat{u} \, , \quad
\hat{q}_5^2 & = y_1 z_1 \hat{s} + x_1 y_1 \hat{t} + x_1 z_1 \hat{u} \, , \nonumber \\
\hat{g}_1^2 & = x_1 y_2 \hat{s} + x_2 y_1 \hat{u} \, , \phantom{+ x_1 y_1 \hat{t}} \quad
\hat{g}_2^2 & = x_2 y_1 \hat{s} + x_1 y_2 \hat{u} \, . \phantom{+ x_1 y_1 \hat{t}} 
\nonumber \\
\label{prop}
\end{eqnarray}
As already indicated in Eq.~(\ref{Tfg}) propagator singularities are treated by
means of the usual $i \epsilon$ prescription. The resulting imaginary part in the amplitudes is
considered as a non-trivial prediction of perturbative QCD which is unaffected by long-distance
effects~\cite{FSZ89}. As we will see below, it gives rise to non-vanishing predictions for 
polarization observables which require the flip of a hadronic helicity. For analytical expressions 
of the 
functions $f$ and $g$ and their symmetry properties we refer to Sec.~\ref{sec:helamp}.

\subsection{Numerical treatment of propagator singularities}
\label{sec:propnum}
The numerical difficulties in performing the convolution integral,
Eq.~(\ref{convol}), are mainly caused by the occurrence of propagator singularities in the range 
of the 
$(x_1,y_1,z_1)$-integration. These give rise to principal value integrals
\begin{equation}
\frac{1}{k^2 \pm i \epsilon} = \wp \left(\frac{1}{k^2}\right) \pm i \pi \delta
(k^2) \, .
\label{pv}
\end{equation}
As explained in some detail in Ref.~\cite{KSPS97} such integrals can still be treated by means 
of a
rather fast and stable fixed-point Gaussian quadrature after carefully separating the principal 
value parts
and doing the corresponding integrations analytically. Here we only want to give a short 
account of our
integration procedure. We first perform the $x_1$-integration for which the various 
contributions to the hard
scattering amplitude (cf. Eq.~(\ref{Tfg})) lead to integrals of the general form
\begin{equation}
I^{(k_1,k_2)} (y_1, z_1) = \int_0^1 dx_1 \frac{h(x_1,y_1,z_1)}{(k_1^2 +
i\epsilon)
(k_2^2 + i\epsilon^{\prime})} \, .
\label{convol2a}
\end{equation}
In order to simplify notations we have neglected helicity labels and the
dependence on the Mandelstam variables $\hat{s}$ and $\hat{t}$. Furthermore, the
distribution amplitudes have been absorbed into the function $h(x_1,y_1,z_1)$.
By applying a partial fractioning, exploiting Eq.~(\ref{pv}), and separating the principal value 
part
Eq.~(\ref{convol2a}) may be rewritten to give
\widetext
\begin{eqnarray}
I^{(k_1,k_2)} (y_1, z_1)  =  \frac{1}{x_1^{(k_1)}-x_1^{(k_2)} +
i\tilde{\epsilon}}
& \Biggl\{  & \left( \frac{\partial k_2^2}{\partial x_1} \right)^{-1}
\int_0^1 dx_1 \frac{h(x_1,y_1,z_1) - h(x_1^{(k_1)},y_1,z_1)}{k_1^2}  \nonumber
\\
  & + & \left( \frac{\partial k_2^2}{\partial x_1} \right)^{-1}
h(x_1^{(k_1)},y_1,z_1) \left( \wp \int_0^1 \frac{dx_1}{k_1^2} - i \pi
\left\vert
\frac{\partial k_1^2}{\partial x_1} \right\vert^{-1} \right) \nonumber \\
& - & \hbox{same}\, (k_1 \leftrightarrow k_2)   \Biggr\} \,
. 
\label{convol3a}
\end{eqnarray}
\narrowtext\noindent
$x_1^{(k_1)} = x_1^{(k_1)}(y_1,z_1)$ and $x_1^{(k_2)} = x_1^{(k_2)}(y_1,z_1)$ are the 
zeroes of the
propagator denominators $k_1^2$ and $k_2^2$, respectively, and $\tilde{\epsilon}$ stands for 
either $\epsilon$
or $\epsilon^{\prime}$, depending for which of the two the limit $\epsilon^{(\prime)} 
\rightarrow 0$ has been
taken. As long as the two zeroes $x_1^{(k_1)}$ and $x_1^{(k_2)}$ (which still depend on 
$y_1$ and $z1$) do
not coincide in the $(y_1,z_1)$-integration region -- this is the case for the contributions of 
$f^{(q,D)}$,
$g^{(q,D)}$, $f^{(D)}$, and $g^{(D)}$ -- the $(x_1,y_1,z_1)$-integration can be performed 
very efficiently
with the help of Eq.~(\ref{convol3a}). After analytic integration of the two principal-value 
integrals in
Eq.~(\ref{convol3a}) (which is a simple matter) all the integrands of the remaining integrations 
are regular
functions of the integration variables and can be treated by fixed-point Gaussian quadrature. 

For $f^{(Q,D)}$ and $g^{(Q,D)}$, however, the situation becomes a little bit more 
complicated. After having
performed the $x_1$-integration in Eq.~(\ref{convol3a}) one observes that the two propagator 
singularities
$x_1^{(k_1)}$ and $x_1^{(k_2)}$ coincide along a trajectory in the $(y_1,z_1)$-integration 
region. This
leads to a quadratic zero $y_1^{(k_1,k_2)} = y_1^{(k_1,k_2)}(z_1)$ of $(x_1^{(k_1)}-
x_1^{(k_2)})$. 
A closer inspection of Eq.~(\ref{convol3a}) reveals that this quadratic zero in the denominator 
of
$I^{(k_1,k_2)} (y_1, z_1)$ does not cause any further problem when performing the
$(y_1,z_1)$-integration for the real part of $I^{(k_1,k_2)} (y_1, z_1)$. The reason is that it is
completely compensated by a corresponding quadratic zero in the numerator of $\Re 
I^{(k_1,k_2)} (y_1,
z_1)$.   In $\Im I^{(k_1,k_2)} (y_1, z_1)$, however,
the zero of $(x_1^{(k_1)}-x_1^{(k_2)})$ is only partly compensated by the numerator and one 
still encounters a
single pole. At the pole $\Im I^{(k_1,k_2)} (y_1, z_1)$ becomes proportional to
$(y_1-y_1^{(k_1,k_2)})/[(y_1-y_1^{(k_1,k_2)})^2+ i \tilde{\epsilon}] = \{1/(y_1-
y_1^{(k_1,k_2)} + i
\hat{\epsilon}) + 1/(y_1-y_1^{(k_1,k_2)}- i \hat{\epsilon})\}/2$. As a consequence $\int_0^1 
dy_1 \Im
I^{(k_1,k_2)} (y_1,z_1)$ becomes a pure principal-value integral (cf. Eq.~(\ref{pv}))
\begin{equation}
\int_0^1 dy_1 \Im I^{(k_1,k_2)} (y_1,z_1) = \wp \int_0^1 dy_1 \Im I^{(k_1,k_2)}
(y_1,z_1)
\label{convol3c}
\end{equation}
which can be treated numerically analogous to the principal-value integral in $x_1$. One only 
has to
separate the principal-value part and do the corresponding integration analytically. For this 
purpose the
analytical expression of the residue of $\Im I^{(k_1,k_2)} (y_1,z_1)$ at the position of the pole
$y_1^{(k_1,k_2)}(z_1)$ is needed. The remaining integrations are then again amenable to 
Gaussian quadrature.

Thus it is possible, by carefully separating the singular contributions, exploiting $\delta$ 
functions,
rewriting principal-value integrals as ordinary integrals plus analytically solvable principal-value
integrals, to do all the numerical integrations by means of fixed-point Gaussian quadrature. 
Whereas
the authors of Ref.~\cite{CRO97}, who have used a different integration procedure, were only 
able to
obtain reliable results for the imaginary part of the convolution integral, Eq.~(\ref{convol}), our 
numerical
results for the imaginary {\bf and} the real part occur to be very stable. Going from an $x_1 
\times
y_1 \times z_1$ grid of $24 \times 24 \times 20$ points to a $48 \times 48 \times 32$ grid 
changes the results
for the amplitudes (over the whole angular range) by less than 0.1\%. Our numerical 
calculations were
performed on an AlphaServer 1000 4/266. For the larger grid size the evaluation of 
$d\sigma/dt(\gamma p
\rightarrow \Phi p)$ (taking into account the 8 hadronic non- and single-flip amplitudes) took 
less than 7
seconds per energy point and angle.

\subsection{Treatment of mass effects}
\label{sec:mass}
Our calculation of the hard-scattering amplitude involves an expansion
in powers of $(m_H/\sqrt{\hat{s}})$ which is performed at fixed scattering angle. We keep 
only
the leading order and next-to-leading order terms in this expansion. Hadron masses, however, 
are
fully taken into account in flux and phase-space factors. The reasons why we want to include 
hadron-mass
effects in the hard scattering amplitude are twofold. On the one hand, we want to apply our 
model already at
momentum transfers of only a few GeV, where hadron masses may still play a role. On the 
other hand, we also
want to make predictions for spin observables which require the flip of hadronic helicities. 
Within 
usual perturbative QCD such observables would vanish since a spin-1 gauge boson which 
couples to a
nearly massless (current) quark cannot flip the helicity of the quark. In the original diquark
model~\cite{Ja93} violations of hadronic helicity conservation are generated by massive V-
diquarks on the
account of introducing a mass parameter for the V-diquark. In this paper we adopt another 
strategy which
allows for a more consistent treatment of mass effects without introducing new mass parameters 
for the
hadronic constituents. Like the authors of Ref.~\cite{ACM90} we assume for every hadronic 
constituent that
its 4-momentum is proportional to the 4-momentum of its parent hadron. This requires to assign 
to every
constituent of the hadron $H$ an effective mass $x m_H$, where $x$ is the fraction of the four-
momentum of
the hadron $H$ carried by the constituent. Keeping in mind that the momentum fractions are 
weighted by the
hadron DAs in the convolution integral, Eq.~(\ref{convol}), this means that quarks and 
diquarks acquire (on
the average) a mass which is rather the mass of a constituent than a current particle. This may be
interpreted in such a way that at intermediate momentum transfers a quark or diquark 
surrounded by a cloud of
$q\bar{q}$ pairs and gluons is not resolved but rather acts like a single effective particle with a
corresponding mass. 

This prescription for the constituent masses applies to the (on-shell) particles which correspond
to the external legs of the Feynman diagrams. It has already been tacitly exploited
when casting the hadronic spin wave functions into covariant form (cf. Eqs.~(\ref{wfcov}) and
(\ref{wfcovv})). Taking the quark-diquark DA of Eq.~(\ref{DAp}) the average values of the 
effective masses
of quarks and diquarks in a proton become
\begin{eqnarray}
\langle m_q^{\text{eff}} \rangle &=& \langle x_1 \rangle m_p \approx \frac{1}{3} m_p \, , 
\nonumber \\
\langle m_D^{\text{eff}} \rangle &=& \langle (1 - x_1) \rangle m_p \approx \frac{2}{3} m_p \, 
.
\label{effmass}
\end{eqnarray}
Analogously the effective mass of the heavy (anti)quark in the meson becomes on the average 
half of the
meson mass. The choice of the masses which occur in the propagators of internal particles is not 
that obvious,
since, e.g., a quark in the incoming proton has mass $x_1 m_p$ whereas the same quark in the 
outgoing proton
has mass $y_1 m_p$. So what should we take for the mass of this quark between two vertices? 
In order to
explain our recipe we, e.g., consider the momentum of the heavy quark between the two gluon 
vertices in the
third diagram of Fig.~\ref{feyn}. It is
\begin{equation}
q_5 = z_1 k + y_1 p_f - x_1 p_i \, . \label{propmom}
\end{equation}
with $p_i$, $p_f$ and $k$ representing the 4- momenta of the hadrons as indicated in 
Fig.~\ref{kinem}.
The quark mass in the propagator related to $q_5$ is now obtained by replacing the 4-momenta 
in
Eq.~(\ref{propmom}) by the masses of the corresponding particles, i.e.
\begin{equation}
m_Q^{(q_5)} =  z_1 m_M + ( y_1 - x_1 ) m_p \, . \label{propmass}
\end{equation}
Taking again into account that the momentum fractions are weighted by the hadron DAs, 
Eq.~(\ref{DAp})
and Eq.~(\ref{DABC}), the average heavy quark mass in the propagator becomes $\langle 
m_Q^{(q_5)} \rangle
\approx 0.5\, m_M$. Extending this recipe to all the quark and diquark propagators and using 
the covariant
wave-function representations, Eqs.~(\ref{wfcov}) and (\ref{wfcovv}), for the external 
particles, leads
to analytical expressions for the Feynman diagrams  which contain (apart of the momentum 
fractions $x_i$,
$y_i$, and $z_i$) only hadronic momenta, masses, spinors, polarization vectors, and helicities. 
The
occurring traces and spinor products can be evaluated in the usual way. The final step is then to 
reexpress
the (massive) Mandelstam variables $s$, $t$, and $u$ in terms of massless ones
\begin{eqnarray}
s & = & \hat{s} (1 + {\cal O}(m_H^2/\hat{s})) \quad \hbox{with} \quad \hat{s} = 4 q^2 \, 
,\nonumber \\
t & = & \hat{t} (1 + {\cal O}(m_H^2/\hat{s}))  
\quad \hbox{with} \quad \hat{t} = - 2 q^2 (1-\cos(\theta_{cm}))\, ,\nonumber \\
u & = & \hat{u} (1 + {\cal O}(m_H^2/\hat{s})) 
\quad \hbox{with} \quad \hat{u} = - 2 q^2 (1+\cos(\theta_{cm})) \, , \nonumber \\
\end{eqnarray}
and to make an expansion in terms of the (small) parameters $(m_H/\sqrt{\hat{s}})$, keeping 
the scattering
angle $\theta_{cm}$ fixed. In order to facilitate the treatment of propagator singularities we 
retain only the
leading and next-to-leading order term in this expansion. Taking again as an example the 
propagator  related
to the momentum $q_5$ (cf. Eq.~(\ref{propmom})) the propagator denominator becomes
\widetext
\begin{eqnarray}
(q_5^2 - {m_Q^{(q_5)}}^2) & = & y_1 z_1 (s - (m_p+m_M)^2)
                          + x_1 y_1 t + x_1 z_1 (u - (m_p-m_M)^2) \nonumber \\
                      & = & y_1 z_1 \hat{s} (1 + {\cal O}(m_H^2/\hat{s}))
                          + x_1 y_1 \hat{t} (1 + {\cal O}(m_H^2/\hat{s}))
                          + x_1 z_1 \hat{u} (1 + {\cal O}(m_H^2/\hat{s})) \nonumber \\
                      & \approx & \hat{q}_5^2
\, ,
\end{eqnarray}
\narrowtext\noindent
hence the quark propagator is given by 
\begin{equation}
\frac{\not{\!q}_5 + m_Q^{(q_5)}}{\hat{q}_5^2 + i \epsilon} \, .
\end{equation}
I.e., by neglecting mass corrections of ${\cal O}(m_H^2/\hat{s})$ we end up with propagator 
denominators which
still appear like those of massless particles (cf. Eq.~(\ref{prop})). Apart from the 
simplifications in the
treatment of propagator singularities and in the analytical expressions for the hard part of the 
amplitudes 
our approximate treatment of mass effects has a few other nice features. By including mass 
corrections of
${\cal O}(m_H/\sqrt{\hat{s}})$ we obtain non-vanishing results for hadron-helicity-flip 
amplitudes and
thus for polarization observables which require the flip of a hadronic  helicity (see the 
discussion in the
next section). Like the leading-order hadron-helicity-conserving amplitudes also the 
hadron-helicity-flip amplitudes are $U(1)$ gauge invariant with respect to the photon 
and $SU(3)$ gauge invariant
with respect to the gluon. Finally, crossing relations for the hadronic helicity amplitudes are 
fulfilled
up to ${\cal O}(m_H/\sqrt{\hat{s}})$, as has been shown for Compton scattering off baryons 
and its crossed
process $\gamma \gamma \rightarrow B \bar{B}$~\cite{Be97}.

\section{Helicity amplitudes \\ and observables}
\label{sec:helamp}
For exclusive photoproduction of vector mesons $\gamma\, p\, \rightarrow\, M\, p$ one finds, 
altogether, 24
complex helicity amplitudes. By virtue of parity invariance only 12 of these helicity amplitudes
are independent. Following the notation of Ref.~\cite{Pi96} we denote them by
\begin{eqnarray}
H_{1,\lambda_\Phi}&=&M_{\lambda_\Phi,\lambda_f=+1/2,\lambda_\gamma=1,\lambda_i=-
1/2} \, , \nonumber \\
H_{2,\lambda_\Phi}&=&M_{\lambda_\Phi,\lambda_f=+1/2,\lambda_\gamma=1,\lambda_i=+
1/2} \, , \nonumber \\
H_{3,\lambda_\Phi}&=&M_{\lambda_\Phi,\lambda_f=-1/2,\lambda_\gamma=1,\lambda_i=-
1/2} \, , \nonumber \\
H_{4,\lambda_\Phi}&=&M_{\lambda_\Phi,\lambda_f=-
1/2,\lambda_\gamma=1,\lambda_i=+1/2} \, ,
\end{eqnarray}
with $\lambda_\Phi = 0, \pm1$. For our normalization of the helicity amplitudes the unpolarized
differential cross section takes on the form
\begin{equation}
\frac{d\sigma}{dt} = \frac{1}{32 \pi (s - m^2_{\text{p}})^2}
\sum_{\lambda_\Phi=0,\pm 1} \sum_{i=1}^4 \vert H_{i,\lambda_\Phi} \vert^2
\, .
\label{unpolwq}
\end{equation}
As examples for single spin observables, for which our model provides nontrivial predictions, 
we will
consider the beam and target asymmetry. The corresponding expressions are~\cite{Pi96}
\widetext
\begin{eqnarray}
\Sigma_x \frac{d\sigma}{dt} & = & \frac {d\sigma_{\|}}{dt} -
\frac{d\sigma_{\perp}}{dt} \nonumber \\ & = & -
\frac{1}{16 \pi (s - m^2_{\text{p}})^2} \nonumber \\ & \times &
\Re \left( H_{4,1}^\ast H_{1,-1} - H_{4,0}^\ast H_{1,0} + H_{4,-1}^\ast H_{1,1} -
H_{3,1}^\ast H_{2,-1}
+H_{3,0}^\ast H_{2,0} -H_{3,-1}^\ast H_{2,1}
\right) 
\label{gasy}
\end{eqnarray}
for the photon and
\begin{eqnarray}
{\cal T}_y \frac{d\sigma}{dt} & = & -
\frac{1}{16 \pi (s - m^2_{\text{p}})^2} \nonumber \\ & \times & 
\Im \left( H_{4,-1}^\ast H_{3,-1} + H_{4,0}^\ast H_{3,0} + H_{4,1}^\ast H_{3,1} + 
H_{2,-1}^\ast H_{1,-1}
+H_{2,0}^\ast H_{1,0} +H_{2,1}^\ast H_{1,1}
\right) 
\label{pasy}
\end{eqnarray}
for the proton, respectively.
\narrowtext

Within the hard scattering approach the energy dependence of the helicity amplitudes at fixed cm 
angle
and large $s$ is roughly
\begin{eqnarray}
H_{2,0}\, &,& \,  H_{3,0} \, \propto s^{-5/2} \, , \nonumber \\
H_{1,0}\, &,& \,  H_{4,0} \, , \, H_{2,1}\, , \,  H_{3,1}  , \, H_{2,-1}\, , \,  H_{3,-1} \, 
\propto
s^{-3} \, ,
\nonumber \\ 
H_{1,1}\, &,& \,  H_{4,1}  , \, H_{1,-1}\, , \,  H_{4,-1} \, \propto s^{-7/2} \, , 
\label{scaling}
\end{eqnarray}
depending on whether the helicity of the hadronic constituents is conserved or flipped by one or 
two
units.  This scaling behavior is modified by  logarithms due to the running coupling
$\alpha_s$ and, eventually, the evolution of the hadron distribution amplitudes. Further 
deviations are
due to the diquark form factors. For small momentum transfers diquarks appear nearly point-
like and the
decay behavior of the helicity amplitudes is weakened by one power of $s$. The energy 
dependence in
Eqs.~(\ref{scaling}) is only approached at large enough momentum transfer, where the diquark 
form factors
become fully operational. 

For the helicity amplitudes our treatment of mass effects (cf. Sec.~\ref{sec:mass}) leads to an 
expansion in
powers of $(m_H/\sqrt{\hat{s}})^2$. As compared to the hadron-helicity-conserving 
amplitudes $H_{2,0}$ and
$H_{3,0}$, the leading-order term of the  single-flip amplitudes $H_{1,0}$, $H_{4,0}$, 
$H_{2,\pm 1}$, and
$H_{3,\pm 1}$ is suppressed by ${\cal O}(m_H/\sqrt{\hat{s}})$. The leading-order term of 
the double-flip
amplitudes $H_{1,\pm 1}$ and $H_{4,\pm 1}$ is even suppressed by 
$(m_H/\sqrt{\hat{s}})^2$ and thus has the
same order as the first (non-vanishing) mass-correction term of the hadron-helicity-conserving 
amplitudes. To
simplify matters we only consider non- and single-flip amplitudes to leading-order and neglect 
double-flip
amplitudes at all.

In the following we quote the analytical results for the functions $f$ and $g$ which have been 
obtained with
the help of the symbolic computer program \lq\lq Mathematica\rq\rq, in particular with the two
program packages \lq\lq \texttt{FEYNARTS}\rq\rq \cite{Kue90} and \lq\lq 
\texttt{FEYNCALC}\rq\rq
\cite{Me91}.  In case of the hadron-helicity conserving amplitude $H_{3,0}$, i.e. for the 
helicity
combination
$ \lambda_\Phi = 0$, $\lambda_f = - 1/2$, $\lambda_\gamma = +1$,
$\lambda_i = -1/2$, the functions $f$ read:
\widetext
\begin{mathletters}
\label{fS1}
\begin{eqnarray}
f_{\left\{ 0,  - \frac{1 }{2 };+1, - \frac{1 }{2 }\right\}}^{(Q,S)}  & = &  - \frac{6 m_p^2 
}{\kappa \hat{t} } f_{\left\{ 0,  - \frac{1 }{2 };+1, - \frac{1 }{2 }\right\}}^{(Q,V)}  
\nonumber \\ & = &
-  2 V \frac{\hat{u}}{x_1 x_2 y_1 y_2 z_1 z_2 \hat{t}^3} \left\{ ( z_2 - z_1
) \left[ z_1 z_2 \hat{u}^2 ( x_1 ( 1+x_2)  - y_1 ( 1 + y_2) ) \right. \right. \nonumber \\ 
& & \left. + \hat{t} \hat{u} ( x_2 + y_2) (x_1 - z_1) (y_1 z_2 + y_2 z_1) - y_2 z_1^2
\hat{t}^2 (1 + y_2) \right] \nonumber \\ 
& & \left. + x_1 y_2 \hat{t}^2 \left[ 4 z_1^2 (y_1 - z_1) +
y_1 (x_2 + z_2) + z_1 (2 x_1 z_1 - 3 y_1) \right] \right\}\, ,
\\
f_{\left\{ 0,  - \frac{1 }{2 };+1, - \frac{1 }{2 }\right\}}^{(q,S)} & = &   - \frac{6 m_p^2 
}{\kappa \hat{t} } f_{\left\{ 0,  - \frac{1 }{2 };+1, - \frac{1 }{2 }\right\}}^{(q,V)} = - 4 V 
\frac{1 }{x_2 y_1 \hat{t} } \left\{ x_2 z_1 \hat{u} + y_2 z_2 \hat{s} \right\}\, ,  \\
f_{\left\{ 0,  - \frac{1 }{2 };+1, - \frac{1 }{2 }\right\}}^{(S)} & = &  - 4 V \frac{1 }{x_1 y_1 
\hat{t} } \left\{ x_1 z_1 \hat{u} + y_1 z_2 \hat{s} \right\}\, , 
\end{eqnarray}
\end{mathletters}
\narrowtext
\noindent
with 
\begin{equation}
V = 32 \pi^2 \sqrt{\pi \alpha}\, \alpha_s^2(\hat{t} \hat{u} / \hat{s}) \sqrt{-\hat{t}} \, C_F \, ,
\label{ofac}
\end{equation}
where $C_F = 2/3 \sqrt{3}$ denotes the color factor and $\alpha$ the fine-structure constant,
respectively. The functions $g$ are related to the $f$s in a simple way: 
\begin{mathletters}
\label{gS1}
\begin{eqnarray}
g_{\left\{ 0,  - \frac{1 }{2 };+1, - \frac{1 }{2 }\right\}}^{(Q,D)} & = & f_{\left\{ 0,  - 
\frac{1 }{2 };+1, - \frac{1 }{2 }\right\}}^{(Q,D)} \quad \, \,\left(z_1 \leftrightarrow z_2 
\right)\, ,\\
g_{\left\{ 0,  - \frac{1 }{2 };+1, - \frac{1 }{2 }\right\}}^{(q,D)} & = & - f_{\left\{ 0,  - 
\frac{1 }{2 };+1, - \frac{1 }{2 }\right\}}^{(q,D)} \,\,\,\left(z_1 \leftrightarrow z_2 \right)\,
,\\ 
g_{\left\{ 0,  - \frac{1 }{2 };+1, - \frac{1 }{2 }\right\}}^{(S)} & = & -f_{\left\{ 0,  - 
\frac{1 }{2 };+1, - \frac{1 }{2 }\right\}}^{(S)} \,\,\,\left(z_1 \leftrightarrow z_2 \right) \, .
\end{eqnarray} 
\end{mathletters}
Vector 4-point contributions $f^{(V)}$ and $g^{(V)}$ are, in general (for all helicity 
combinations),
suppressed by ${\cal O}(m_H^2/\hat{s})$ or even stronger. Relations analogous to 
\hbox{Eqs.~(\ref{gS1})}
\hbox{are} also valid between the $f$s and $g$s entering the other $\lambda_\Phi = 0$ 
amplitudes $H_{1,0}$,
$H_{2,0}$, and $H_{4,0}$. The functions $f$ contributing to $H_{2,0}$ --
the other hadron-helicity conserving amplitude (helicity combination $
\lambda_\Phi = 0$, $\lambda_f = + 1/2$, $\lambda_\gamma = +1$, $\lambda_i = + 1/2$) -- are 
obtained from
those entering $H_{3,0}$ by interchange of the Mandelstam variables
$\hat{s} \leftrightarrow \hat{u}$ and the momentum fractions $x_1 \leftrightarrow y_1$:
\begin{mathletters}
\label{fS2}
\begin{eqnarray}
f_{\left\{ 0,  + \frac{1 }{2 };+1, + \frac{1 }{2 }\right\}}^{(Q,D)} & = & f_{\left\{ 0,  - 
\frac{1 }{2 };+1, - \frac{1 }{2 }\right\}}^{(Q,D)}  \quad \, \left( \hat{s} \leftrightarrow 
\hat{u}, x_1 \leftrightarrow y_1 \right) \, ,\nonumber \\ & & \\
f_{\left\{ 0,  + \frac{1 }{2 };+1, + \frac{1 }{2 }\right\}}^{(q,D)} & = & - f_{\left\{ 0,  - 
\frac{1 }{2 };+1, - \frac{1 }{2 }\right\}}^{(q,D)} \,\, \left( \hat{s} \leftrightarrow \hat{u}, x_1 
\leftrightarrow y_1 \right) \, ,\nonumber \\ & & \\
f_{\left\{ 0,  + \frac{1 }{2 };+1, + \frac{1 }{2 }\right\}}^{(S)} & = &  -f_{\left\{ 0,  - \frac{1 
}{2 };+1, - \frac{1 }{2 }\right\}}^{(S)} \,\, \left( \hat{s} \leftrightarrow \hat{u}, x_1 
\leftrightarrow y_1 \right) \, .\nonumber \\ & &
\end{eqnarray} 
\end{mathletters}
The analytical expressions for the hadron-helicity-flip amplitudes $H_{1,0}$, $H_{4,0}$, 
$H_{2,\pm 1}$,
and $H_{3,\pm 1}$ are of similar length as the expressions quoted above and can be obtained 
from the
authors on request. Like in Eqs.~(\ref{fS1}) the vector 3-point contributions to $H_{2,\pm 1}$ 
and $H_{3,\pm
1}$ are proportional to the corresponding scalar ones. It is also interesting to observe that due to 
the
(anti)symmetry properties of the Feynman graphs under interchange of the momentum fractions 
$z_1$ and $z_2$
(cf. Eqs.~(\ref{gS1})), only diagrams in which the photon couples to the $s$ quark contribute 
to the
$\lambda_\Phi = 0$ amplitudes. 

Actually we have computed all the functions $f$ and $g$, and relations like (\ref{gS1}) and 
(\ref{fS2})
served as a check of our analytical expressions. Furthermore, we have verified $U(1)$ gauge
invariance with respect to the photon and $SU(3)$ gauge invariance with respect to the gluon. 
The proof
of gauge invariance is facilitated by the observation that not only the sum of all 16 tree diagrams
provides a gauge invariant expression, but rather each of the functions $f$ and $g$ is gauge 
invariant by
itself. Finally, we have recalculated a few of the diagrams by hand to confirm the outcome of 
Mathematica.
\section{Numerical results}
\label{sec:result}
Fig.~\ref{result} shows the diquark model predictions (solid line) for $d\sigma/dt [ \gamma p 
\rightarrow
\Phi p ]$ at $p_{\text{lab}}^{\gamma}=6$~GeV together with the outcome of the pomeron-
exchange mechanism
(dotted line)~\cite{DoLa87} and results of a two-gluon-exchange model (dash-dotted and 
dashed
line)~\cite{La95,La98}. A direct comparison of our predictions with experiment is not possible 
yet since data
are only available at low momentum transfers ($t\, \alt\, 1.5$~GeV$^2$) where the perturbative
photoproduction mechanism is certainly not the dominant one. The experimental situation, 
however, will
hopefully improve as soon as the data analysis of the CEBAF-93-031 experiment will be 
completed.
Nevertheless, Fig.~\ref{result} shows that the diquark model provides a reasonable 
extrapolation of the
low-$t$ data. Thereby one should keep in mind that the cross section which decays 
exponentially in forward
direction is expected to flatten around $90^{\circ}$ ($t$ and $u$ large). The few available data 
for
photoproduction of pseudoscalar mesons and also photoproduction of $\rho$ and $\omega$ 
mesons which extend to
larger values of $t$ exhibit just such a behavior~\cite{An76}. As can be seen from 
Fig.~\ref{result} the
forward cross section for $\phi$ production is reasonably well reproduced by simple Pomeron 
phenomenology
(dotted line). As a mechanism for photoproduction of vector mesons Donnachie and 
Landshoff~\cite{DoLa87}
proposed that the photon fluctuates into a quark-antiquark pair which subsequently recombines 
to a vector
meson. The resulting quark loop is connected to a quark in the proton via the exchange of a 
single pomeron
which couples to quarks like an isoscalar photon. The QCD-inspired version of the Pomeron 
exchange of Donnachie
and Landshoff~\cite{DoLa89}, in which the Pomeron is replaced by two non-perturbative 
(abelian) gluons, has
been applied by Laget and Mendez-Galain~\cite{La95} to photoproduction of $\Phi$ mesons 
(dash-dotted line).
The latter authors consider the two-gluon-exchange model as a link which connects the 
diffractive with the
hard-scattering domain. At $t \approx 2.5$~GeV$^2$ the two-gluon-exchange cross section 
exhibits a
characteristic node which can be understood as an interference effect between the two Feynman 
diagrams which
enter the photoproduction amplitude. However, this picture changes drastically if the two 
gluons are allowed
to couple to different quarks in the proton. Recently Laget~\cite{La98} tried to estimate such 
contributions
by modeling quark-quark correlations inside a proton via a simple wave function. He found that 
the node in
the cross section is completely washed out and that the contributions in which the two gluons 
couple to
different quarks inside the proton start to dominate the photoproduction of $\Phi$ with 
increasing $t$
(dashed line). Actually, for $\vert t \vert \agt 4$~GeV$^2$ the result of Laget becomes 
comparable with the
diquark-model result. This is not surprising since we know from the perturbative analysis that 
the
hard-scattering mechanism, i.e. contributions from diagrams without loops in which all the 
hadronic
constituents are connected via gluon exchange, should become dominant for large values of
$t$ and $u$. Correlations between hadronic constituents are automatically accounted for within 
such an
approach and hence also within our diquark model. 

To check whether our model provides a reasonable energy dependence we have also calculated 
the differential
cross section at a much larger photon energy ($p_{\text{lab}}^{\gamma}=2611$~GeV). The 
result is depicted in
Fig~\ref{result2} (solid line) along with a two-parameter fit of $\vert t \vert \leq 0.5$~GeV$^2$ 
data from
ZEUS~\cite{ZEUS96}. The hatched area indicates the uncertainties in the forward cross section 
and the slope
parameter. Keeping again in mind that the cross section is expected to flatten with increasing $t$ 
the
magnitude of our prediction occurs still to be within the range of an extrapolation of the low-$t$ 
data. The
dashed line in Fig.~\ref{result2} represents a leading-order calculation in which mass effects are 
neglected.
This means that  only the two hadron-helicity conserving amplitudes $H_{2,0}$ and 
$H_{3,0}$ are taken into
account in the leading-order cross section. A comparison between the full calculation (solid line) 
and the
leading-order calculation (dashed line) reveals that mass effects are still sizable and amount to 
$\approx
25$-$40\%$ of the full differential cross section. They increase in forward and backward 
direction and are
smallest around $90^{\circ}$. 

At this point we want to comment on another attempt to calculate photoproduction of $\Phi$ 
mesons within
the diquark model. The authors of Ref.~\cite{CRO97} use a different parameterization of the 
model. Only the
hadron-helicity-conserving amplitudes $H_{2,0}$ and $H_{3,0}$ are taken into account in 
their calculation.
With their numerical method they encounter difficulties in calculating the real parts of these 
amplitudes.
But they assert that the real parts are small for their parameterization and therefore neglect them. 
Their
results for differential cross sections are about one order of magnitude smaller than ours. This is 
plausible
if one looks at their values of $f_S^p$ and $f_V^p$, i.e. the \lq\lq normalization\rq\rq\ of the
quark-diquark DAs, and the cutoff masses in the diquark form factors. Both are considerably 
smaller than
ours. In contrast to the authors of Ref.~\cite{CRO97} we think that the present experimental 
situation in
$\Phi$ photoproduction does not  allow to discriminate a particular parameterization of the 
diquark model.
However, we want to point out that our parameterization has already been applied successfully 
to a variety
of other exclusive processes~\cite{Ja93,Kro93,Kro95,Be97,KSPS97}. On the other hand, it is 
still unclear
whether these processes can also be described satisfactorily with the parameterization of 
Carimalo et
al.~\cite{CRO97}. Their parameterization has been determined by means of the proton magnetic 
form factor.
Apart from $\Phi$ photoproduction it has only been applied to charmonium decays into the $p\, 
\bar{p}\,
\gamma$ final state~\cite{CO91} with the result that the data could not be reproduced.

In the kinematic region, where $t$ and $u$ are large, the differential cross section at fixed 
scattering
angle is expected to exhibit a scaling behavior of approximately $s^{-7}$, which is supposed to 
indicate that the
hard photoproduction mechanism becomes relevant. Fig.~\ref{scalingfig} shows $s^7\, 
d\sigma / dt$ for photon
energies of $p_{\text{lab}}^{\gamma}=5$, $10$, and $20$~GeV. 5 GeV is presently the 
upper limit of CEBAF,
10 GeV will be reached by the planned upgrade of CEBAF and 20 GeV is a value which could 
be reached at future
facilities like ELFE or a further extension of CEBAF which are presently still under discussion. 
Comparing
the diquark-model results for these energies one observes that one is still far away from the 
$s^{-7}$
scaling behavior. The reasons have already been mentioned in Sec. \ref{sec:helamp}. A closer 
inspection
shows that the cross section at $\theta_{\text{cm}} = 90^{\circ}$ scales rather like
$s^{-8}$ than $s^{-7}$ for $10 < s < 40$~GeV.  Starting with point-like diquarks and a 
constant 
$\alpha_{\text{s}}$ one would expect that the leading-order cross section (without mass 
effects) should scale
like $s^{-5}$. By including the diquark form factors and cross-section contributions from
hadron-helicity-flip amplitudes, which are already suppressed by one power of $s$ as compared 
to the
leading-order contributions,  the scaling behavior changes already to nearly $s^{-7}$. Finally, 
the
approximate $s^{-8}$ decay is caused by the fact that $\alpha_{\text{s}}$ is taken as a running 
coupling
constant (with argument $(\hat{t} \hat{u} / \hat{s})$). 

Fig.~\ref{effects} shows how the differential cross section (at 
$p_{\text{lab}}^{\gamma}=6$~GeV) is influenced
by different contributions within our model. A comparison of the solid (full calculation) and the
short-dashed (mass effects neglected) line reveals, e.g., the influence of mass effects. For
$p_{\text{lab}}^{\gamma}=6$~GeV and $\cos (\theta_{\text{cm}})=90^{\circ}$ they amount 
to $\approx 40\%$ of
the full cross section. The dash-dotted line represents the pure S-diquark part of the cross 
section. It
varies between $15\%$ and $30\%$ if one goes from backward to forward angles. Since S- and 
V-diquark
contributions add coherently (and constructively, cf. Eq.~(\ref{fS1})) this means that the 
corresponding
amplitudes are of the same order of magnitude. The differential cross section for the 
photoproduction of
longitudinally polarized $\Phi$s is given by the long-dashed curve. It amounts to $\approx 
60\%$ of the
full cross section at $\cos (\theta_{\text{cm}})=90^{\circ}$. This fraction increases in 
backward direction
and decreases in forward direction. Finally, we have also examined the influence of the 
exponential in the
$\Phi$ DA (cf. Eq.~(\ref{DABC})). Neglecting this exponential one ends up with the dotted 
curve which is
nearly indistinguishable from the full calculation. This also shows that we are not very sensitive 
to the
$z_1 \rightarrow 0,1$ endpoint regions when performing the convolution integral 
Eq.~(\ref{convol}).

As examples for single-polarization observables the beam asymmetry $\Sigma_x$ and the target 
asymmetry ${\cal
T}_y$ are depicted in Fig.~\ref{asymmetry}. The beam asymmetry would be already non-zero, 
if only the
leading-order helicity amplitudes $H_{2,0}$ and $H_{3,0}$ were taken into account (cf. 
Eq.~(\ref{gasy})). It
would not even be necessary that the helicity amplitudes acquire a non-trivial phase. On the 
other hand, a
non-trivial target asymmetry demands for both, non-vanishing hadron-helicity-flip amplitudes 
and non-trivial
phases of the amplitudes (cf. Eq.~(\ref{pasy})). As already mentioned at the end of 
Sec.~\ref{sec:elamp}
phases are generated through propagator singularities,  hadron-helicity-flip amplitudes occur in 
our model as
mass corrections. As it turns out within our model, also the beam asymmetry is considerably 
affected by the
inclusion of the hadron-helicity-flip amplitudes. In a recent paper~\cite{ZDGS99} the influence 
of nucleonic
resonance effects on polarization observables in $\Phi$-meson photoproduction has been 
investigated within a
constituent-quark model in which the diffractive contribution is produced by t-channel Pomeron 
exchange. It
is interesting to observe that the result of these authors for the target asymmetry at a rather small 
photon 
energy of $p_{\text{lab}}^{\gamma}=2$~GeV is comparable in size to our result at 
$p_{\text{lab}}^{\gamma}=6$~GeV. This is
remarkable since two completely different concepts are at work: on one hand nucleonic 
resonance
contributions, on the other hand a purely perturbative mechanism. Also for the beam asymmetry 
some
similarities can be found. The corresponding predictions of both approaches are again 
comparable around
$\theta_{\text{cm}} = 90^{\circ}$. A (somewhat smaller) negative beam asymmetry has also 
been found in
another model which is based on pomeron exchange, pion exchange, and a direct knockout 
production from the
strange-quark sea in the proton~\cite{TOYM98}. It would be interesting to see experimentally, 
whether
polarization observables like the target asymmetry ${\cal T}_y$, the recoil polarization ${\cal
P}_{y^{\prime}}$, or the vector-meson polarization ${\cal V}_{y^{\prime}}$, which vanish 
in leading-order
perturbative QCD, are still sizable at momentum transfers of a few GeV. Such observables 
would be a more
sensitive tool than the unpolarized differential cross section to figure out whether the kind
of nonperturbative ingredients (diquarks, effective constituent masses) which are taken into 
account in our
perturbative approach, are already sufficient, or whether much more non-perturbative physics is 
still at
work at a few GeV of momentum transfer.

At the end of this section we want to comment on the perspective to extend this calculation to
photoproduction of $J/\Psi$. It is a simple matter, since we have already assumed a flavor 
dependent DA for
the vector meson (cf. Eq.~(\ref{DABC})). The only thing to do is to replace the $\Phi$ mass by 
the $J/\Psi$
mass and the decay constant $f^{\Phi}$ by $f^{J/\Psi}$. One, however, has to keep in mind 
that we have applied
an expansion in the parameter $(m_H/\sqrt{\hat{s}})$. In order that this parameter is 
approximately
of the same size in $\Phi$ and $J/\Psi$ production the photon lab energy has to be increased 
approximately by
a factor $(m_{J/\Psi}/m_{\Phi})^2 = 9.2$ when going from $\Phi$ to $J/\Psi$. A second 
criterion for the
reliability of our approximate treatment of mass effects is that the coefficient of the first mass 
correction
term should become small compared to the leading order. This coefficient is angular dependent 
and becomes
large at small scattering angle. That means, if the size of the mass corrections relative to the 
leading order
should be about the same in $\Phi$ and $J/\Psi$ production ($(m_H/\sqrt{\hat{s}})$ 
comparable), one has to
consider much larger momentum transfers for $J/\Psi$ than for $\Phi$ production. A more 
detailed discussion
of this point along with some numerical results can be found in Ref.~\cite{BS99}. A possible 
way to apply
the perturbative approach to $J/\Psi$ production at a few GeV of momentum transfer would be 
to retain our
prescription for the choice of the effective masses, but to refrain from the final expansion in
$(m_H/\sqrt{\hat{s}})$. This has not been attempted yet since the analytical expressions for the
amplitudes become very lengthy and the treatment of propagator singularities in the convolution 
integral,
Eq.~(\ref{convol}), becomes much more involved.

\section{Summary and conclusions}
\label{sec:sum}
We have investigated the reaction $\gamma\, p\, \rightarrow\, \Phi\, p$ in the few-GeV 
momentum transfer region
where unpolarized differential cross-section data from the CEBAF-93-031 experiment at JLAB 
are expected to
become available very soon. Our theoretical approach is based on perturbative QCD 
supplemented by the
assumption that baryons can be treated as quark-diquark systems. The same approach has 
already been
successfully applied to other photon-induced hadronic 
reactions~\cite{Ja93,Kro93,Kro95,Be97,KSPS97}. The
motivation for introducing diquarks is, above all, to extend the perturbative approach from large 
down to
moderately large momentum transfers. There are mainly two reasons why this can be achieved 
with diquarks.
First, the overall momentum transfer has to be shared between less hadronic constituents than in 
the pure quark HSA. Hence the gluons which keep the hadronic constituents collinear are on the 
average harder and thus
the running strong coupling constant becomes smaller. The second reason is that a certain class 
of
non-perturbative effects, namely quark-quark correlations inside a baryon, are effectively 
accounted for by
means of diquarks. The fact that a heavy quarkonium is produced in the final state entails a 
considerable
reduction in computational effort. Only a small fraction of the Feynman diagrams which 
contribute in general
to photoproduction of arbitrary mesons occurs in $\Phi$ production. To be more specific, these 
are just those
diagrams in which the photon fluctuates into the heavy quark-antiquark pair which is then 
connected to the
quark and the diquark in the proton via the exchange of two gluons, respectively. Both, the 
scalar and the
vector-diquark component of the proton have been taken into account. We have also improved  
the  diquark
model in the sense that hadron-mass effects in the hard-scattering amplitude have been included 
in
a systematic way. This has been accomplished by assigning to each hadronic constituent an 
effective mass  $x
m_H$ ($x \dots$ longitudinal momentum fraction of the constituent) and expanding the 
scattering amplitude in
the small parameter $(m_H/\sqrt{\hat{s}})$. Only the leading and next-to-leading order terms 
in this
expansion have been kept. The leading-order term provides for the hadron-helicity-conserving 
amplitudes and
agrees, of course, with the massless result. The next-to-leading order term represents a power-
correction and
gives rise to non-vanishing hadron-helicity-flip amplitudes. It enables us to make predictions for
polarization observables which require the flip of a hadronic helicity like, e.g., the target 
asymmetry or the 
recoil polarization. We want to point out that our treatment of mass effects still preserves gauge 
invariance 
with respect to the photon and the gluon. 

Our numerical studies have been performed with the diquark-model parameters and the quark-
diquark DAs
proposed in Ref.~\cite{Ja93}. The DA of the $\Phi$ meson has been taken from the 
literature~\cite{BC90}. It
fulfills QCD sum-rule constraints. Hence we have been able to make predictions without 
introducing new
parameters.  We have paid special attention to the correct and numerically robust treatment of 
propagator
singularities. In the absence of large-$t$ data ($t \agt 4$~GeV$^2$) our results can only be
compared with the available data which go up to $t \approx 1.5$~GeV$^2$. Our predictions for 
the
differential cross section at two rather different photon energies 
($p_{\text{lab}}^{\gamma}=6$~GeV and
$2600$~GeV) which are typical for CEBAF and HERA, respectively, occur to be still within 
the range of an
extrapolation of the existing low-$t$ data. It is worth mentioning that the predictions of a
two-gluon-exchange model, in which the gluons are considered as a remnant of the 
pomeron~\cite{La98}, become
comparable to our result for $t \agt 4$~GeV$^2$ if the gluons are allowed to couple to different 
quarks in the
proton. This is not surprising because contributions in which all the hadronic constituents are
(approximately) kept collinear by means of gluon exchange to prevent the hadrons from 
breaking up become
increasingly important as soon as one enters the hard-scattering domain. Just those 
contributions are assumed
to be the dominant ones within our HSA-based diquark model. We have also investigated 
within our model
whether the (asymptotic) $s^{-7}$ scaling behavior of the differential cross section at fixed 
angle is already
realized for $s$ values between $10$ and $40$~GeV. After dividing out the strong running 
coupling (which
occurs to the sixth power in the cross section) an approximate $s^{-7}$ scaling has been found.  
Our
investigation of hadron-mass effects has revealed that they are  still non-negligible in the few-
GeV momentum
transfer range and photon energies of the order of 10~GeV. Correspondingly, also polarization 
observables
which require the flip of a hadronic helicity, like the target polarization, occur to be sizable. 
Finally,
we have considered the possibility to extend the present calculation to photoproduction of 
$J/\Psi$
mesons. In order that our treatment of mass effects remains reliable for $J/\Psi$s one has to go 
to much
larger energies and momentum transfers where experimental data cannot be expected. The 
situation with respect
to the momentum transfer could perhaps be improved by refraining from the final expansion in
$(m_H/\sqrt{\hat{s}})$. This, however, will considerably complicate the treatment of 
propagator singularities
and make the analytical expressions for the amplitudes much more complicated. Therefore it has 
not been
attempted as yet.

Our diquark-model calculations for $\gamma p \rightarrow \Phi p$ will meet the first severe test 
in the near
future when the data analysis of the CEBAF 93-031 experiment will be completed. With these 
data, which
go up to $t$ values of about 5~GeV$^2$, one will get a first glimpse how different production 
mechanisms
(pomeron exchange $\rightarrow$ gluon exchange $\rightarrow$ hard-scattering mechanism) in 
high-energy
photoproduction evolve with increasing momentum transfer. However, in order to really touch 
the
hard-scattering domain much higher photon energies than those which can be reached at the 
present stage of CEBAF
would be desirable. In addition to unpolarized differential cross sections, polarization 
observables which 
require the flip of hadronic helicities, could serve as an indicator for the relative importance of 
hard and
soft photoproduction mechanisms, since hard scattering is closely connected with hadronic 
helicity
conservation. 

\acknowledgments
C. F. Berger would like to thank the Paul-Urban-Stipendienstiftung for supporting a visit at the
Institute of Theoretical Physics at the University of Graz during which part of this paper has 
been
completed.

\begin{figure}
\caption{Graphical representation of the  hard-scattering formula, Eq.~(\ref{convol}),
for $\gamma \, p\, \rightarrow\, M \, p$. $x_i$, $y_i$, and $z_i$ denote
longitudinal momentum fractions of the constituents.}
\label{kinem}
\end{figure}

\begin{figure}
\caption{Eight of the, altogether, sixteen tree-level diagrams which contribute to
the elementary process $\gamma\,q\, D\,  \rightarrow\, Q \, \bar{Q}\, q \, D$. The remaining
diagrams are obtained by interchanging the two gluons. We have also indicated how the 
diagrams
are grouped to gauge invariant expressions.}
\label{feyn}
\end{figure}

\begin{figure}
\caption{Differential cross section for $\gamma\, p\, \rightarrow\, \Phi\, p$ versus $t$ at
$p_{\text{lab}}^{\gamma}=6$~GeV. The solid line corresponds to  the diquark-model 
prediction. Cross sections
resulting from the Pomeron-exchange mechanism~{\protect \cite{DoLa87}} (dotted line) and a 
two-gluon-exchange
model~{\protect \cite{La95}} (dash-dotted line) are plotted for comparison. The dashed line 
represents an
attempt to extend the two-gluon-exchange model by including contributions in which the two 
gluons couple to
different quarks inside the proton~{\protect \cite{La98}}. Data are taken from Ballam et 
al.~{\protect
\cite{Bal73}} and Barber et al.~{\protect
\cite{Ba83}}.}
\label{result}
\end{figure}

\begin{figure}
\caption{Differential cross section for $\gamma\, p\, \rightarrow\, \Phi\, p$  versus $t$ at
$p_{\text{lab}}^{\gamma}=2611$~GeV ($s = 4900$~GeV$^2$). The diquark-model 
prediction with (solid line) and
without (dashed line) mass effects is compared with  a parameterization of the low-$t$ ZEUS-
data~{\protect
\cite{ZEUS96}}. The hatched area indicates the uncertainties in the forward cross section and 
the slope
parameter.}
\label{result2}
\end{figure}

\begin{figure}
\caption{Differential cross section for $\gamma\, p\, \rightarrow\, \Phi\, p$ scaled by $s^7$  
versus $\cos
(\theta_{\text{cm}})$. Diquark-model results for $p_{\text{lab}}^{\gamma}=5$ (solid line), 
$10$ 
(long-dashed line), and $20$~GeV (dashed line).}
\label{scalingfig}
\end{figure}

\begin{figure}
\caption{Differential cross section for $\gamma\, p\, \rightarrow\, \Phi\, p$ scaled by $s^7$  
versus $\cos
(\theta_{\text{cm}})$  at $p_{\text{lab}}^{\gamma}=6$~GeV. A comparison of different 
contributions within the
diquark model: full calculation (solid line), mass effects neglected (short-dashed line), S 
diquarks only
(dash-dotted line), longitudinally polarized $\Phi$s only (long-dashed line), exponential in 
$\Phi$-DA
neglected (dotted line).}
\label{effects}
\end{figure}

\begin{figure}
\caption{Diquark-model predictions for beam ($\Sigma_x$) and target (${\cal T}_y$) 
asymmetry in 
$\gamma\, p\, \rightarrow\, \Phi\, p$ scattering at $p_{\text{lab}}^{\gamma}=6$~GeV.}
\label{asymmetry}
\end{figure}


\newpage

\begin{figure}
\epsfig{file=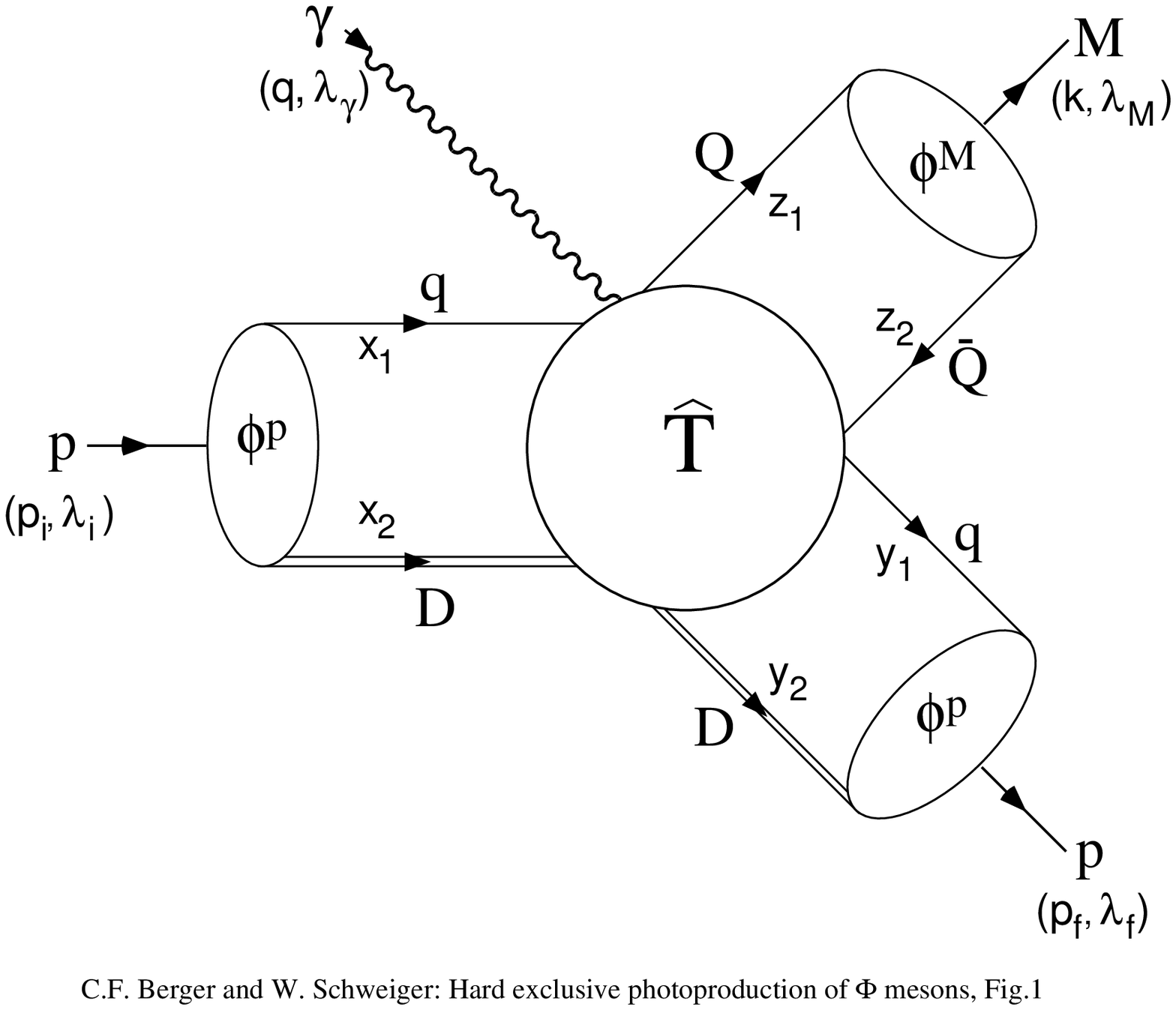,width=12.0cm,clip=}
\end{figure}

\begin{figure}
\epsfig{file=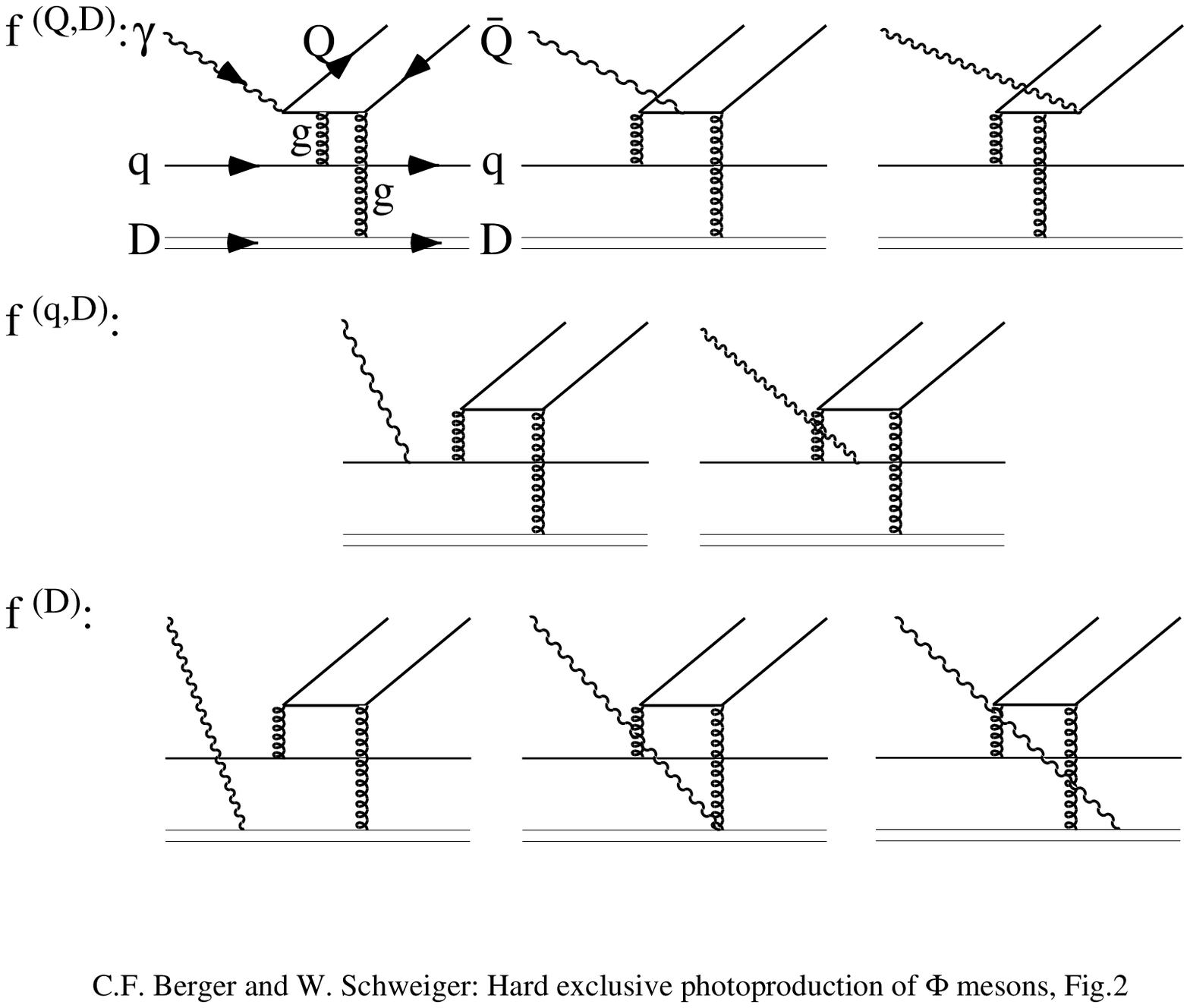,width=12.0cm,clip=}
\end{figure}

\newpage

\begin{figure}
\epsfig{file=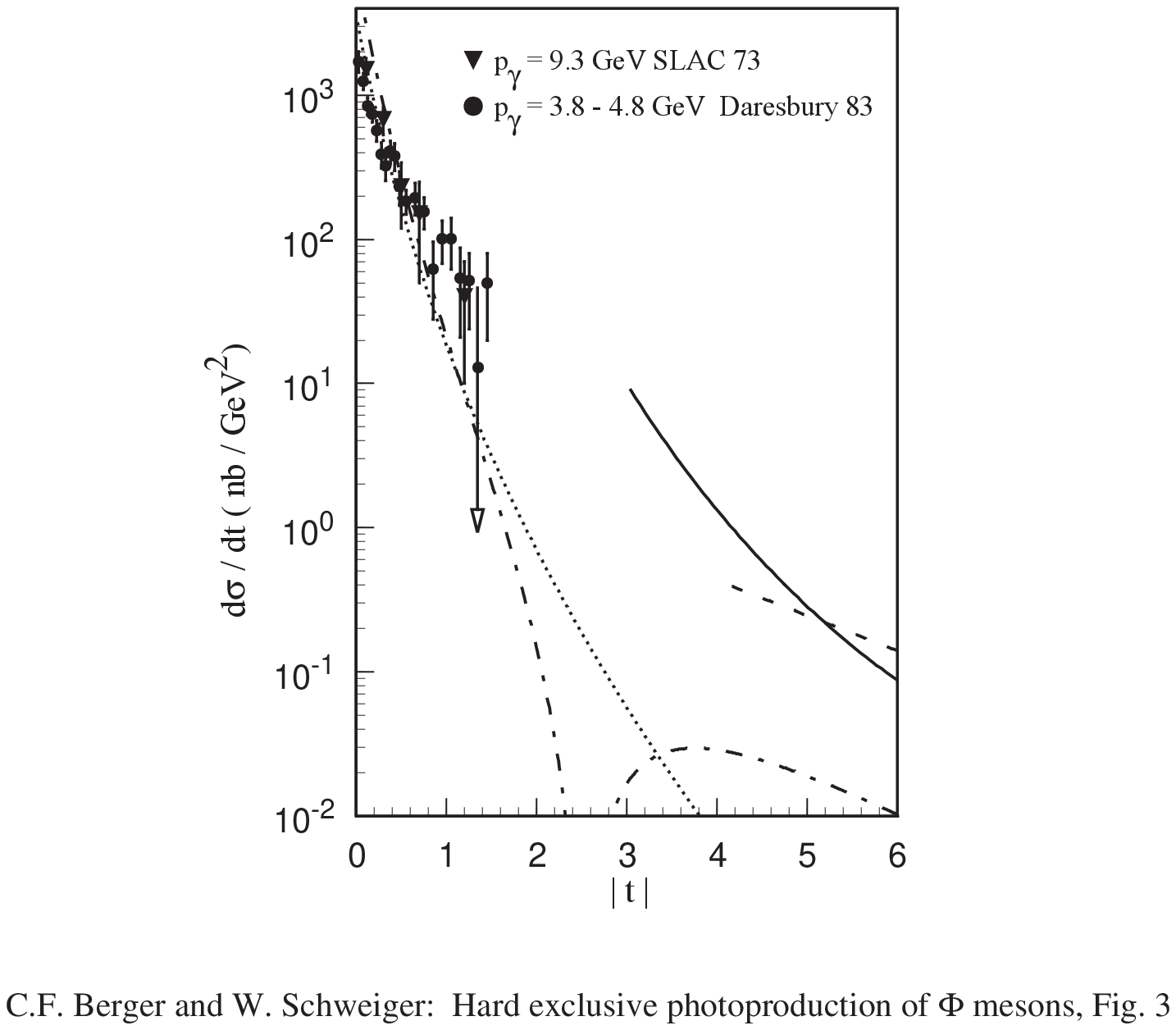,width=12.0cm,clip=}
\end{figure}

\begin{figure}
\epsfig{file=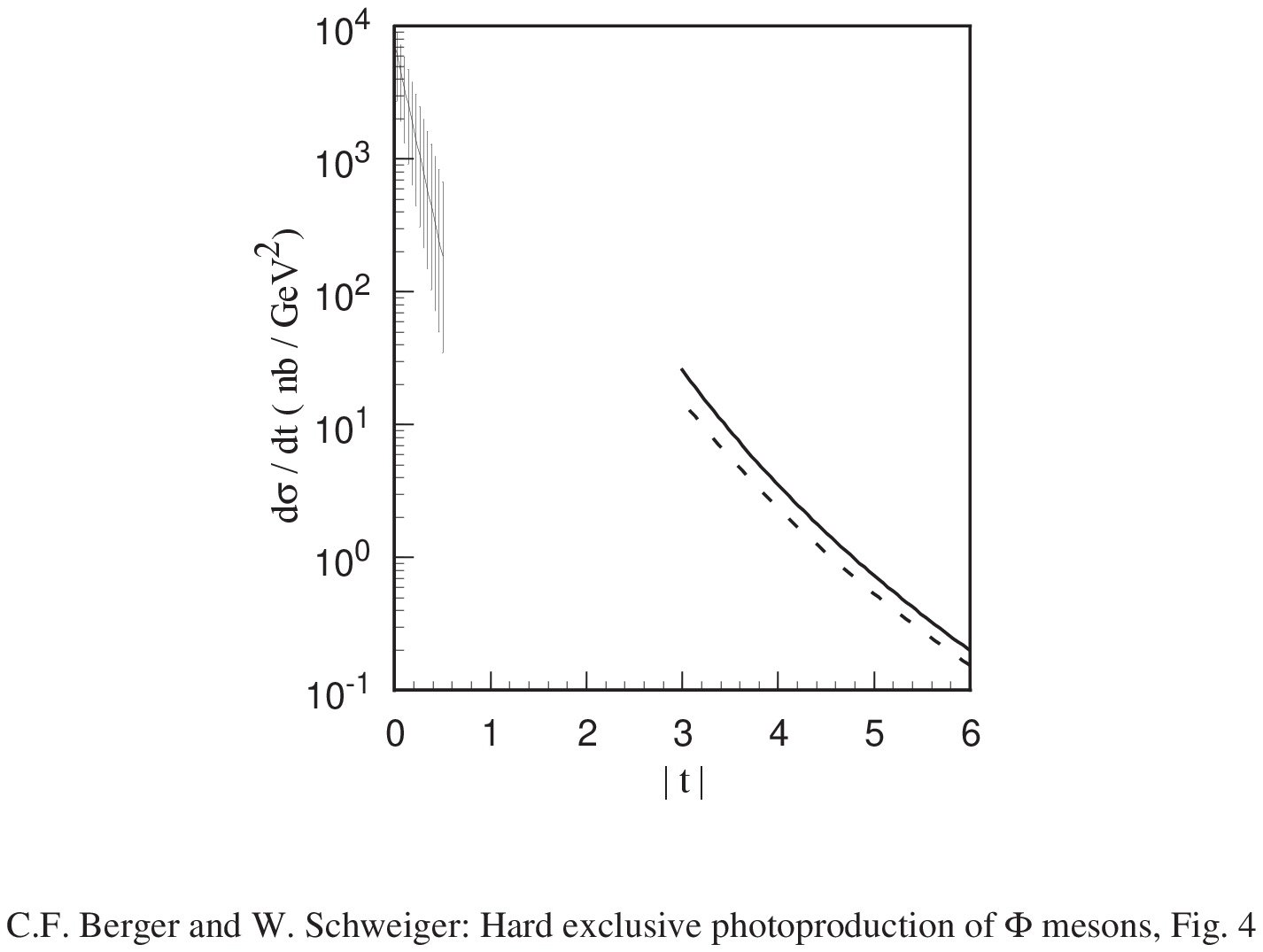,width=12.0cm,clip=}
\end{figure}

\newpage

\begin{figure}
\epsfig{file=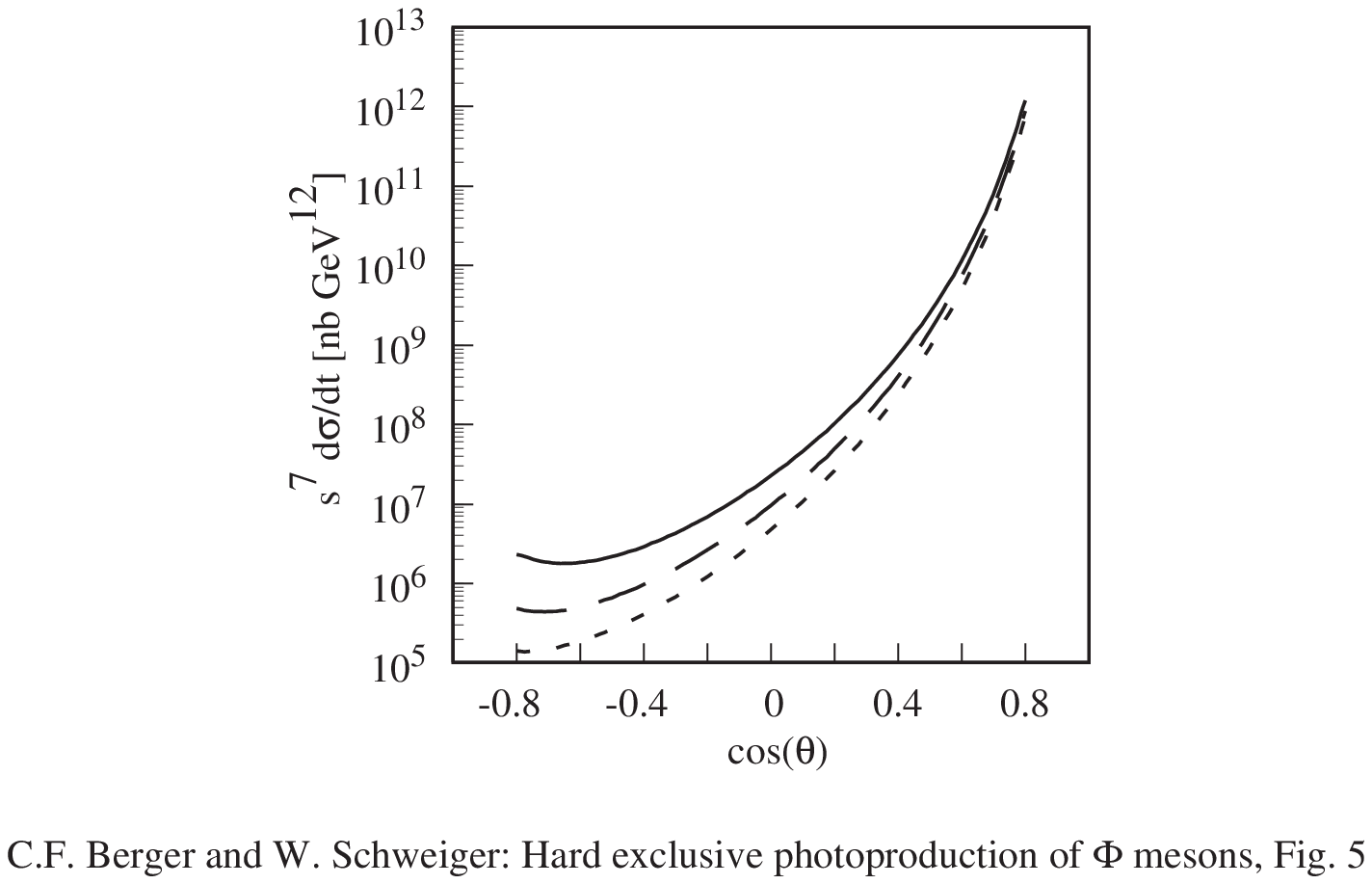,width=12.0cm,clip=}
\end{figure}

\begin{figure}
\epsfig{file=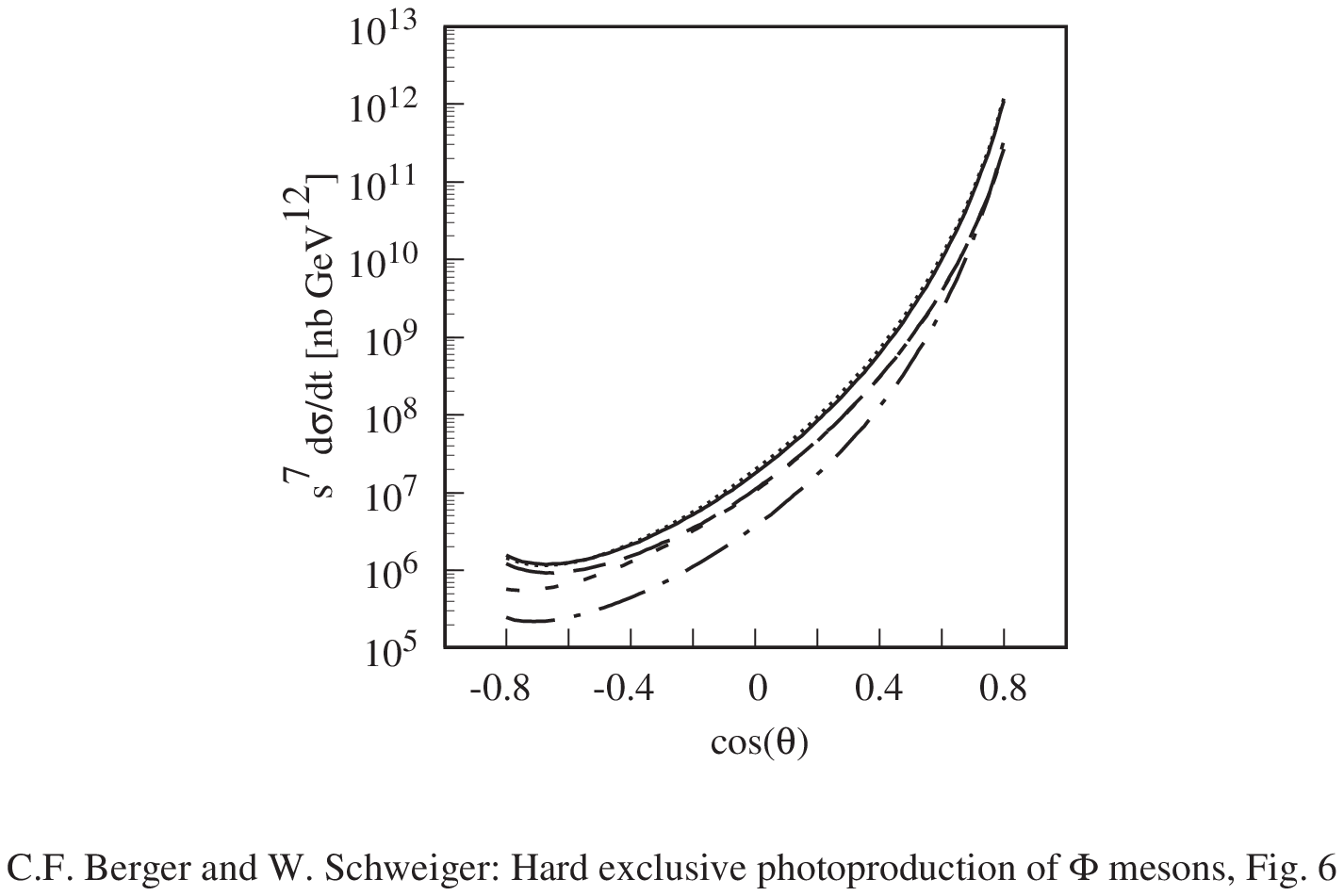,width=12.0cm,clip=}
\end{figure}

\newpage

\begin{figure}
\epsfig{file=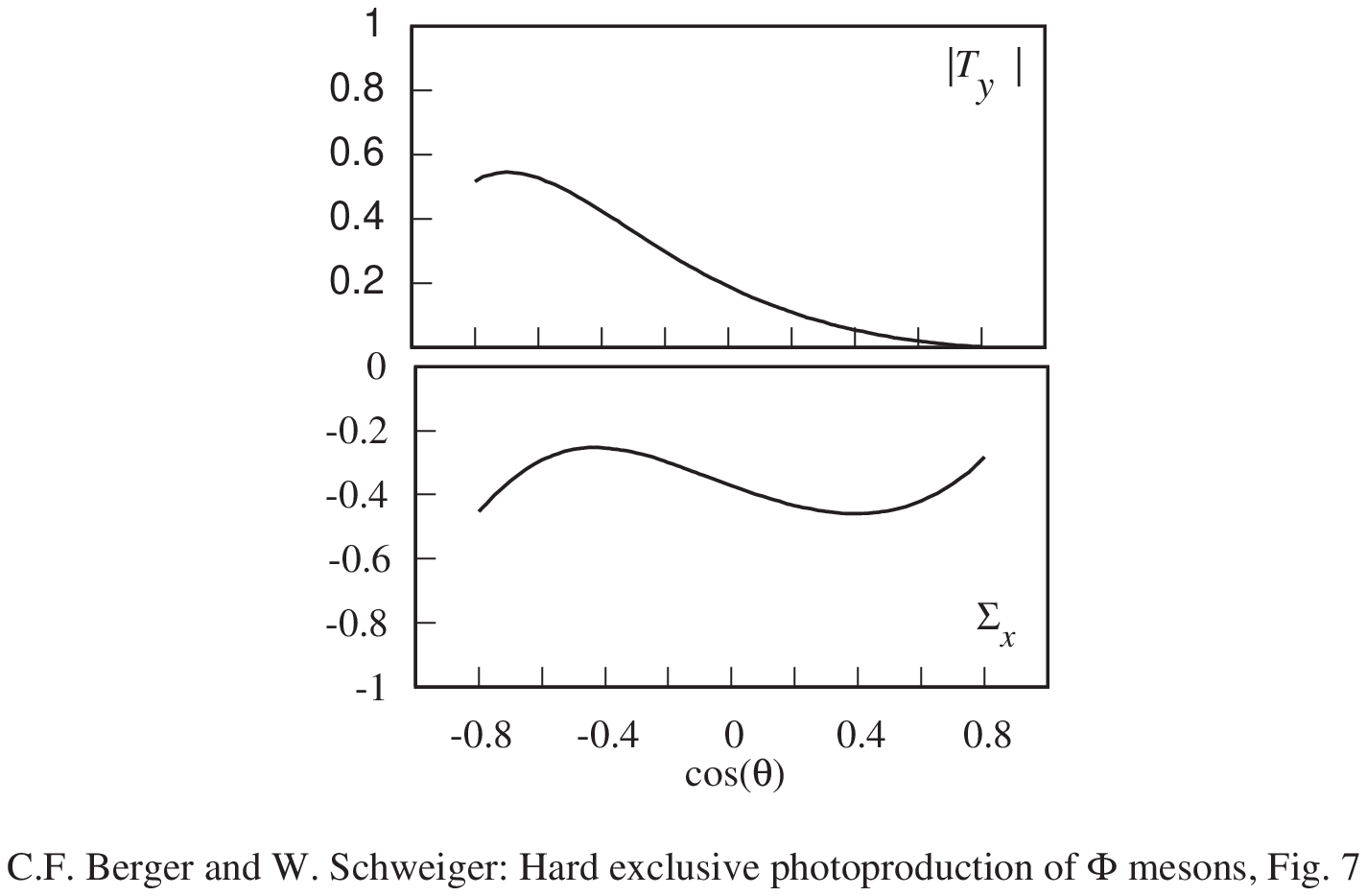,width=12.0cm,clip=}
\end{figure}

\end{document}